\documentclass[11pt,a4paper]{article}
\pdfoutput=1
\usepackage{jcappub}
\usepackage[english]{babel}

\definecolor{nicered}{rgb}{0.7,0.1,0.1}
\definecolor{nicegreen}{rgb}{0.1,0.5,0.1}
\hypersetup{colorlinks,citecolor= nicegreen,linkcolor= nicered}

\newcommand{\be}{\begin{equation}}
\newcommand{\ee}{\end{equation}}
\newcommand{\bea}{\begin{eqnarray}}
\newcommand{\eea}{\end{eqnarray}}
\definecolor{Red}{rgb}{1.,0.,0.}

\def\ede{\end{equation}}
\def\bga{\begin{aligned}}
\def\eda{\end{aligned}}
\newcommand{\beq}{\begin{equation}}
\newcommand{\eeq}{\end{equation}}
\newcommand{\bq}{\begin{equation}}
\newcommand{\eq}{\end{equation}}
\newcommand{\ba}{\begin{array}}
\newcommand{\ea}{\end{array}}
\newcommand{\beqa}{\begin{eqnarray}}
\newcommand{\eeqa}{\end{eqnarray}}
\newcommand{\beqs}{\begin{subequations}}
\newcommand{\eeqs}{\end{subequations}}

\def\nn{\nonumber}

\def\dis{\displaystyle}

\def\({\left(}
\def\){\right)}

\def\RB{\right]}

\def\leqq{\leqslant}

\def\End{\end{document}}

\def\d{\text{d}}

\def\over{\overline}

\def\be{\beta}

\def\PhiN{\Phi_{\text{N}}^{}}
\def\PhiDE{\Phi_{\text{DE}}^{}}
\def\N{\text{N}}
\def\DE{\text{DE}}
\def\M{\text{M}}

\def\rr{\mathbf{r}}
\def\xx{\mathbf{x}}
\def\xxc{\mathbf{x}_c^{}}
\def\CC{\Lambda}
\def\O{\cal O}
\def\S{\cal S}
\def\rc{r_{\text{cri}}^{}}
\def\reff{r_{\!\text{eff}}^{}}
\def\rff{r_{\!\text{eff}}^2}
\def\rfff{r_{\!\text{eff}}^3}
\def\neff{n_{\text{eff}}^{}}
\def\ro{r_{\!o}^{}}
\def\B{\beta}
\def\rrr{r^3}
\def\rccc{r_{\text{cri}}^3}
\def\RRR{r^3\!/r_{\text{cri}}^3}
\def\rrreff{r_{\text{eff}}^3}
\def\RB{\over{R}}
\def\rb{\bar{r}}

\def\rh{\hat{r}}

\def\End{\end{document}}

\title{\huge Direct Probe of Dark Energy through Gravitational Lensing Effect}

\author[a,b]{\large Hong-Jian He,}
\author[b]{\large~~Zhen Zhang\,}

\affiliation[a\,]{Institute of Modern Physics and Center for High Energy Physics,\\
                  Tsinghua University, Beijing 100084, China.}
\affiliation[b\,]{Center for High Energy Physics, Peking University, Beijing 100871, China.}
\emailAdd{hjhe@tsinghua.edu.cn, zh.zhang@pku.edu.cn}

\abstract{
\\[1mm]
We show that gravitational lensing can provide a direct method to probe the nature
of dark energy at astrophysical scales. For lensing system as an isolated astrophysical object,
we derive the dark energy contribution to gravitational potential as a repulsive power-law term,
containing a generic equation of state parameter $w$\,.
We find that it generates $w$-dependent and position-dependent modification
to the conventional light orbital equation of $\,w=-1$\,.
With post-Newtonian approximation, we compute
its direct effect for an isolated lensing system at astrophysical scales and find that
the dark energy force can deflect the path of incident light rays.
We demonstrate that the dark-energy-induced deflection angle
$\,\Delta\alpha_{\DE}^{}\propto M^{(1+\frac{1}{3w})}\,$
(with $1\!+\!\frac{1}{3w}>0$), which increases with the lensing mass $M$ and
consistently approaches zero in the limit $M\!\!\to\!0$\,.\,
This effect is distinctive because dark energy tends to diffuse the rays and generates
{\it concave lensing effect.}
This is in contrast to the conventional convex lensing effect caused by both visible
and dark matter. Measuring such concave lensing effect can directly probe the existence
and nature of dark energy. We estimate this effect and show that the current gravitational
lensing experiments are sensitive to the direct probe of dark energy at astrophysical scales.
For the special case $\,w=-1$\,,\, our independent study favors the previous works that
the cosmological constant can affect light bending, but our predictions qualitatively
and quantitatively differ from the literature, including our consistent realization
of $\,\Delta\alpha_{\DE}^{}\!\to\!0$ (under $M\!\!\to\!0$) at the leading order.
}

\keywords{\\[1mm]
Dark Energy Theory, Gravitational Lensing, 
Particle Physics$-$Astrophysics Connection
\\[4mm]
~\hfill JCAP (2017) [arXiv:1701.03418 [astro-ph.CO]].
}

\begin{document}

\maketitle

\setlength{\baselineskip}{18pt}

\setcounter{page}{2}
\vspace*{10mm}
\section{\hspace*{-1.5mm}Introduction}
\label{sec:1}
\vspace*{2mm}

The discovery of cosmic acceleration has pointed to the mysterious
dark energy, which composes about 69\% of the total energy density of the
present universe \cite{Planck2015} and
remains one of the greatest puzzles of modern science.
To unravel the mystery of dark energy faces two challenges:
one is to find out direct observational evidence
of dark energy\,\cite{DEexp}, and another is to identify
the right theory for describing the dark energy\,\cite{DE-Rev}.

The current dark energy detections\,\cite{DEexp} are mostly indirect,
taking the form of large astronomical surveys and including such as
the CMB measurements of WMAP and Planck,
the Type-Ia Supernovae, the Baryon Acoustic Oscillations,
the Gravitational Lensing, the Clusters of Galaxies, and so on.
One of these approaches is the gravitational lensing\,\cite{Bartelmann99},
which has become an important method to probe both the dark matter
and dark energy\,\cite{Heavens09,GL-Rev}
since its observation in 1979 \cite{Walsh79}.
So far the gravitational lensing analyses are through the
indirect probe of dark energy,
where the dark energy effect is indirectly included via the scale factor
$a(t)$ defined 
under Friedmann-Lemaitre-Robertson-Walker (FLRW) metric
and the matter contribution is treated as perturbation.
Because the CMB measurements\,\cite{Planck2015} at cosmological scale
show that the FLRW spacetime is nearly flat
(with spatial curvature $K\simeq 0$) and the matter perturbation
only generates Newtonian deflection, the conventional lensing analyses
find no measurable direct effect of dark energy, as is obvious.
But, for most of the gravitational lensings at the (shorter) astrophysical scales
(such as the scales of galaxies and galaxy clusters), the cosmological expansion
factor $a(t)$ is nearly constant,
and the dark energy effect should be best included
via its {\it direct contribution} to the gravitational potential
as a repulsive force {\it in parallel to} the matter contribution.
This can be derived under the Schwarzschild-de\,Sitter spacetime
with a generic equation of state parameter $w$\, (denoted as SdS$w$ metric),
and we solve Einstein equation for the isolated lensing system in the SdS$w$ spacetime
(Appendix\,\ref{app:0}).

In this work, we study the direct probe of dark energy through
the gravitational lensing at astrophysical scales.
We will establish a setup to define the astrophysical-scale
gravitational lens as an isolated system with an effective radius $\,\reff\,$
(characetrized by its critical radius $\,\rc\,$
at which the dark energy repulsive force just cancels
the attractive Newtonian force of matter).
For such a lensing system, we use the post-Newtonian approximation to
derive the general gravitational potential including
both the Newtonian term and the dark energy term (with a generic equation of state
parameter $w$\,) in the SdS$w$ spacetime.
Inside this system, we demonstrate that the dark energy force directly
contributes to the lensing effect together with the Newtonian force.
For the case of a single lensing,
in the regions outside this isolated lensing system
($\,\rrr \gg \rfff\!\sim\!\rccc\,$),\,
the Newtonian potential becomes negligible and
the spacetime conformally recovers the FLRW metric
with vanishing spatial curvature $K=0$\,,\,
which is conformally flat and thus will not change the deflection angle
of the light ray.

We will further demonstrate that the dark-energy-induced deflection angle
$\,\Delta\alpha_{\DE}^{}\propto M^{(1+\frac{1}{3w})}\,$
(with $1\!+\!\frac{1}{3w}>0$\,), which increases with the lensing mass $M$ and
consistently approaches zero under the limit $\,M\!\!\to\!0$\,.\,
This effect is distinctive because dark energy tends to diffuse the rays and generates
{\it concave lensing effect,}
contrary to the conventional convex lensing effect caused by both visible
and dark matter. Measuring such concave lensing effect can directly probe the existence
and nature of dark energy. We will estimate this effect and show that the current gravitational
lensing experiments are sensitive to the direct probe of dark energy at astrophysical scales.
We also note that
for the special case $w=-1$, our independent study favors the previous works of
Ishak and collaborators\,\cite{Ishak-Rev}\cite{Ishak2007} that
the cosmological constant can affect light bending. But our predictions qualitatively
and quantitatively differ from the literature\,\cite{Ishak-Rev},
including our consistent realization
of $\,\Delta\alpha_{\DE}^{}\!\!\to\!0$ (under $M\!\!\to\!0$) at the leading order.

This work is organized as follows.
In section\,\ref{sec:2}, we introduce the dark energy potential with a generic equation
of state parameter $w$\,,\, for an isolated astrophysical system.
In section\,\ref{sec:3}, we analyze the dark energy lensing effect,
and demonstrate that it acts as a {concave lensing}, in contrast with the
conventional convex lensing effect caused by both visible and dark matter.
In section\,\ref{sec:4}, we estimate the dark energy lensing effect,
and show that the current gravitational lensing experiments are sensitive to
the direct probe of dark energy.
In section\,\ref{sec:5A}, for the lensing system
with a point-like spherical mass-distribution,
we derive a new orbital equation of light rays
for a generic state parameter $w$\,.\,
It generates $w$-dependent and position-dependent modification
to the conventional light orbital equation of $\,w=-1$\,.\,
We will conclude in section\,\ref{sec:5}.
In Appendix\,\ref{app}, we derive the general dark energy potential
used in the main text and discuss its connection with the typical dark energy models.
We also clarify the difference of our independent approach from
the previous studies of the cosmological constant case ($w=-1$)
\cite{Ishak-Rev} at the end of Appendix\,\ref{app:A4}.
In Appendix\,\ref{app:B1}, we derive the light orbital equation in the
SdS$w$ spacetime.
Finally, Appendix\,\ref{app:CC} presents two related analyses within and outside the lensing system.

\vspace*{3mm}
\section{\hspace*{-1.5mm}Gravitational Potential Including Dark Energy}
\label{sec:framework}
\label{sec:2}
\vspace*{1mm}

In the conventional analysis of gravitational lensing,
the gravitational potential is entirely determined by matter
(including visible matter and dark matter).
But we note that dark energy can directly modify the form of
gravitational potential at astrophysical scales.
This gives rise to a correction term $\,\Delta\Phi\,$
in the gravitational potential, and should be included to
directly describe the gravitational lensing effects.
Following Ref.\,\cite{Weinberg72}, we can write the metric of
an isolated astrophysical system, under the post-Newtonian approximation,
\beqa
\d{S}^{2} = (1\!+\!2\Phi)\d t^{2}\!-\! (1\!-\!2\Phi)
(\d\textit{x}^{2}\!+\d\textit{y}^{2}\!+\d\textit{z}^{2}),
\hspace*{7mm}
\label{eq:metric1}
\label{eq:dS2}
\eeqa
where  $\,\Phi =\Phi^{}_{\N}\!+\!\Delta\Phi\,$
is the gravitational potential,
containing the conventional Newtonian potential $\,\Phi^{}_{\N}\,$
and the correction term $\,\Delta\Phi=\Phi_{\DE}^{}\,$
as induced by dark energy. 
For the post-Newtonian region within the effective radius
($\,r < \reff\sim\rc\,$),\, the metric \eqref{eq:metric1} holds well
and the dark energy contributaion $\,\Phi_{\DE}^{}\,$ behaves
as an effective potential (cf.\ Appendix\,\ref{app:11}).
As we show in Eqs.\eqref{eq:PhiNDE} and \eqref{eq:Phi-NDE2}
of Appendix\,\ref{app:11},
for such an isolated astrophysical system, the dark energy contribution
$\,\Phi_{\DE}^{}\,$ to the gravitational potential $\,\Phi\,$ is unavoidable,
and must be included together with the matter contribution $\,\Phi_{\N}\,$.\,
Hence, we can study the direct contribution of $\,\Phi_{\DE}^{}\,$
to the gravitational lensing at astrophysical scales.

In the spherical coordinate system, we locate the center of mass $M$
at the origin and adopt the geometrized unit system ($G=h=c=1$).
Then, we can solve Einstein equation exactly in Appendix\,\ref{app:11}
and obtain the complete potential form,
\beqa
\Phi \,=\, \Phi_{\N}^{} + \Phi_{\DE}^{}
\,=\, -\frac{\,M\,}{r}-\(\!\frac{\,\ro\,}{r}\!\)^{\!\!3w+1},~~~~
\label{potential}
\label{eq:Phi}
\eeqa
where the Newtonian term $\,\Phi_{\N}^{}=-M/r\,$
holds for a point-like mass $M$ or for regions outside a spherically symmetric
mass-distribution. For a general case of a mass-density $\rho(\rr)$
distributed over a space region $\Omega$, the total mass is
$\,M=\int_{\Omega}^{}\!\d^3 \rr'\rho(\rr')$\,,\, and the Newtonian potential is
$\,\Phi_{\N}^{}=-\int_{\Omega}^{}\!\d^3 \rr'{\rho(\rr')}/{|\rr -\rr'|}\,
$.\,
In Eq.\eqref{eq:Phi}, the constant $\,w\,$ is the equation of state parameter
of dark energy and is given by $\,w =p/\rho\,$ as the ratio between the pressure
$\,p\,$ and energy density $\,\rho$\,.\,
The parameter $\,r_o^{}\,$ of $\,\Phi_{\DE}^{}$
characterizes the size of the present universe and
will be given in each dark energy model when comparing
with the cosmological data.

Eq.\eqref{eq:Phi} shows that the dark energy induced correction term
$\,\Phi_{\DE}^{}\,$ takes a model-independent form.
It can describe the equation of state for different dark energy models\,\cite{DEmodel}:
$\,w =-1\,$ for the cosmological constant dominated state,
$\,-1< w <-\frac{1}{3}\,$ for the quitessence dominated state,
and $\,w <-1\,$ for the phantom dominated state.
For the cosmological constant model of dark energy,
the dark energy induced gravitational potential takes the form,
$\,\Phi_{\DE}^{} =-({r}/{\ro})^{2}$\,
with $\,\ro = \sqrt{6/\CC}\,$,\,
which will be discussed in Appendix-\ref{app:3}.

\vspace*{3mm}
\section{\hspace*{-1.5mm}Direct Dark\,Energy\,Effect in Gravitational Lensing}
\label{sec:del}
\label{sec:3}
\vspace*{1mm}

After including the direct correction of dark energy to the Newtonian potential,
the metric takes the form of Eq.\eqref{eq:dS2} in the post-Newtonian region,
where the full gravitational potential
$\,\Phi=\Phi_{\N}^{}+\Phi_{\DE}^{}\,$
contains both the matter and dark energy contributions as given by Eq.\eqref{eq:Phi}.
We consider the potential to be fairly weak,
$\,\Phi_{\N}^{}\ll 1$\, and $\,\Phi_{\DE}^{}\ll\,1$\,.\,
Under this weak field approximation,
the full potential \eqref{eq:Phi} is derived from the Einstein equation
as in Appendix\,\ref{app:11}.

Photons propagate along the light-like path with $\,\d{S}^2=0\,$.\,
Let $\,\d\ell =(\d x,\, \d y,\, \d z)$ denote
the change of spatial coordinates $(x,\,y,\,z)$ of the photon
during a time interval $\,\d t\,$.\,
Thus, under linear approximation, we deduce from Eq.\,\eqref{eq:dS2},
\beqa
\d t \,=\,
\(1\! -2\Phi_\N^{}\!-2\Phi_{\DE}^{}\)
\d\ell\,.
\label{dtdl}
\label{eq:dt-dl}
\eeqa

In the space with gravitational field,
a photon propagates precisely as if there is no gravitational field,
but the space is filled with refractive medium.
From Eq.(\ref{dtdl}), we deduce the refractive index $\,n=n(x,y,z)$\,,
\beqa
\textit{n} \,=\, n_{\M}^{}+\Delta n_{\DE}^{}
\,=\, 1-2\Phi_{\N}^{}-2\Phi_{\DE}^{}\,,
\label{zeshelv}
\eeqa
where $\,n_{\M}^{}\,$ is the conventional matter contribution, and
$\,\Delta n_{\DE}^{}\,$ is the dark energy induced correction term,
\beqa
n_{\M}^{}=1-2\Phi_{\N}^{}\,, ~~~~
\Delta n_{\DE}^{} = -2\Phi_{\DE}^{}\,.
\eeqa
\begin{figure}[t]
\centerline{
\includegraphics[width=14cm,height=7cm]{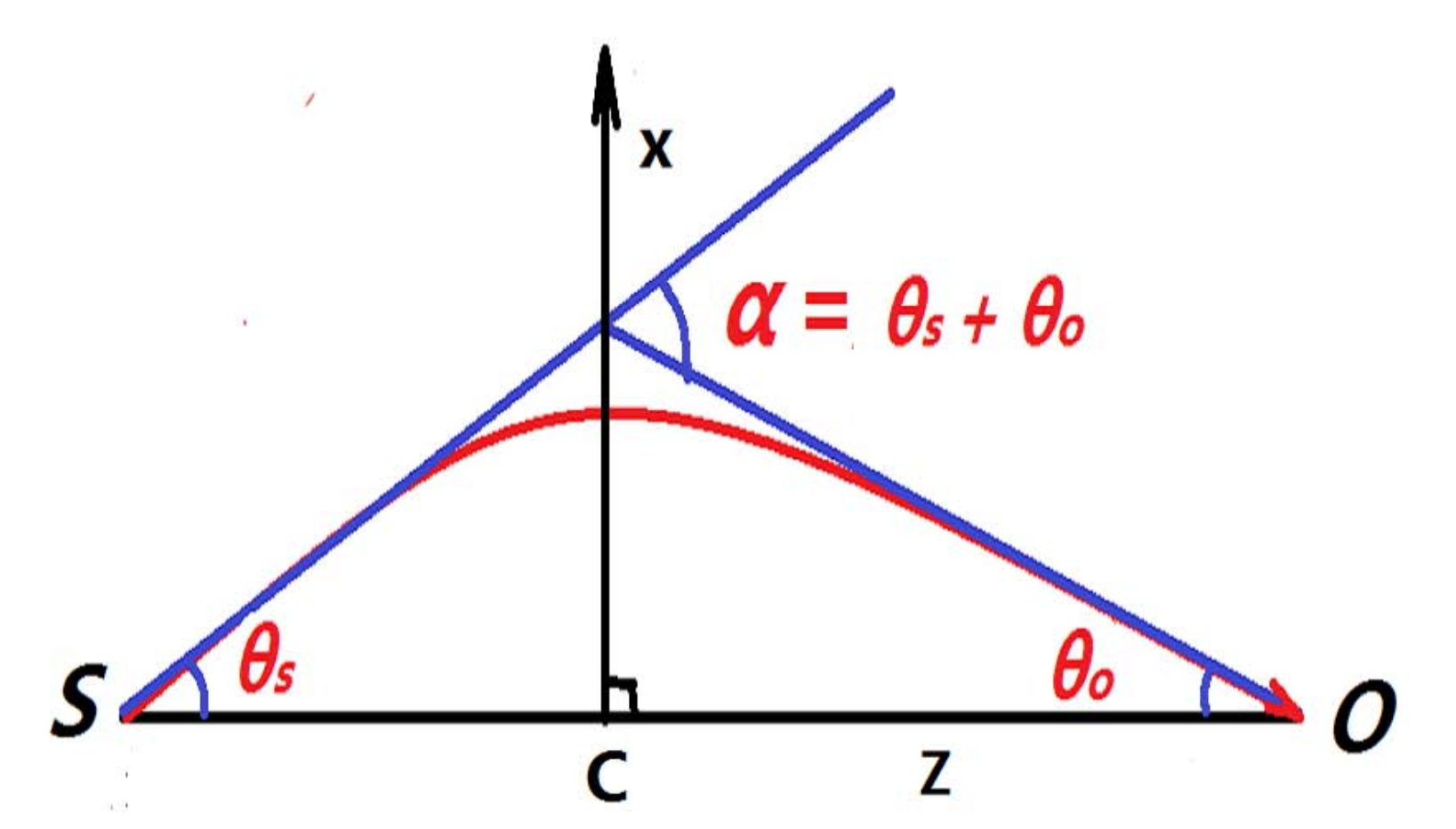}}
\vspace*{-2mm}
\caption{Convex gravitational lensing.
The red curve represents the projected trajectory of a ray
in the $x$-$z$ plane, which is deflected by the gravitational field of matter.
The $\alpha$ denotes the angle between the directions of the ray at the point $\S$
and at the point $\O$. The incident angle at $\S$\, is $\theta_{\S}^{}$,\,
and the outgoing angle at $\O$ is $\theta_{\O}^{}$.\, The center of mass $M$
sits at the point $C$\,.}
\label{fig:DM}
\label{fig:1}
\end{figure}

In Fig.\,\ref{fig:DM}, the red curve shows
that the trajectory of a ray\footnote{Here, a ray generically refers to
any possible ray of electromagnetic wave, such as light ray, $X$ ray, and $\gamma$ ray.
(It could also be certain relativistic particles.)
In the present work, we will often mention the light ray
as a familiar example without losing generality.}
from source to observer is bent.
In the post-Newtonian region,
we locate the source at point $\cal S$ and the observer at point $\cal O$,\,
while the center of mass $M$ sits at the origin $C$.
The ray propagates through the deflecting region between the two points
$\S$ and $\O$\,.\, Under the weak field approximation, it is reasonable to consider
the potential $\,\Phi\ll 1$\,,\, and thus we have $\,n\simeq 1\,$
at the $\S$ and $\O$ points. Let us orient the coordinate system such that
the straight line $\overrightarrow{\S\O}$ is parallel to the $z$-axis, and assign
one of its vertical directions as the $x$-axis.
The Fermat's principle demands\,\cite{Fermat}\cite{Fermat2},
\beqa
\(\!n\frac{\rm{d}x}{\rm{d}\ell}\!\)_{\!\!\O}^{}
-\(\!n\frac{\d x}{\d\ell}\!\)_{\!\!\S}^{} \,=\,
\int_{\S}^{\O}\!\!\d\ell\,\frac{\,\partial n\,}{\partial x} \,.
\label{dangle}
\label{eq:FPpath}
\eeqa
In the post-Newtonian region, we have\footnote{%
For instance, in the relevant lensing region around $\,r=(0.02\!-2)\rc$\,
and for $\,w\simeq -1\,$,\,
we can estimate $\,\Phi\lesssim\,(0.05-1)\!\times\! 10^{-3}\ll 1\,$,\,
for typical galaxies or galaxy clusters with masses $\,M\lesssim 10^{16}M^{}_{\odot}$.}\,
$\Phi\ll 1\,$,\,
and thus $\,n_{\S}^{}\simeq n_{\O}^{}\simeq 1\,$.\,
Thus, the left-hand-side (LHS) of Eq.\eqref{eq:FPpath} can be replaced by the sum of
the two Euclidean geometry angles, $\,-(\theta_{\O}^{\text{E}}\!+\theta_{\S}^{\text{E}})\,$,\,
where $\,\theta_{\S}^{\text{E}}=\!\(\!\frac{\d x}{\d\ell}\!\)_{\!\S}^{}\,$ and
$\,\theta_{\O}^{\text{E}}=-\!\(\!\frac{\d x}{\d\ell}\!\)_{\!\O}^{}\,$.\,
The Euclidean angles are connected to the
measurable angles $(\theta_{\S}^{},\,\theta_{\O}^{})$
via relation \cite{LL},
%
$\tan(\theta_{I}^{})=
\sqrt{1\!+\!2\Phi\,}\,{\tan}(\theta_{I}^{\text{E}})$\,
%
where $I=\S,\O$,\, and $\,\theta_{\S}^{}\,$ ($\,\theta_{\O}^{}\,$) denotes
the physical incident (outgoing) angle as shown in Fig.\,\ref{fig:1}.
Since $\,\Phi\ll 1\,$,\,
we thus expand this relation and deduce
%
$
\theta_{I}^{\text{E}}\simeq \theta_I^{}$\,.\,
%
Then, the LHS of Eq.\eqref{eq:FPpath} can be further expressed as
$\,-(\theta^{\text{E}}_{\O}\!+\theta^{\text{E}}_{\S})\simeq
-(\theta^{}_{\O}\!+\theta^{}_{\S})$.\, Hence, Eq.\eqref{eq:FPpath} becomes 
\beqa
-(\theta^{}_{\O}\!+\theta^{}_{\S})
\,\simeq\,
\dis \(\! n\frac{\d x}{\d\ell}\!\)_{\!\!\O}^{}
    -\(\! n\frac{\d x}{\d\ell}\!\)_{\!\!\S}^{}, ~~~
\label{eq:alpha-1}
\label{twoangle}
\eeqa

The deflection angle $\,\vec{\alpha}\,$ is the difference between
the directions of the incident ray at the point $\S$ and the outgoing ray at the point $\O$\,,\,
$\,\vec\alpha =\vec\theta^{}_{\O}\!-\vec\theta^{}_{\S}\,$,\,
with $(\vec{\theta}_{\S}^{},\,\vec{\theta}_{\O}^{},\,\vec{\alpha})$ defined as
(counterclockwise,\,clockwise,\,clockwise).  
In the $x$-$z$ plane, this gives
$\,\alpha =\theta^{}_{\S}\!+\theta^{}_{\O}\,$ as shown by Fig.\,\ref{fig:1}, where 
$\,(\theta_{\S}^{},\,\theta_{\O}^{},\,\alpha)
=(|\vec{\theta}_{\S}^{}|,\,|\vec{\theta}_{\O}^{}|,\,|\vec{\alpha}|)\,$.\,
The same relation holds for the $y$-$z$ plane.
In general, for any similar plane, by substituting
Eq.(\ref{twoangle}) into Eq.(\ref{dangle}), we have
\begin{equation}
\vec{\alpha}  \,=\, \vec{\theta}_{\O}^{}-\vec{\theta}_{\S}^{}
\,=\, -2\!\int_{\S}^{\O}\!\!\!\d\ell\,\boldsymbol{\nabla}_{\!\!\perp}^{}\!\Phi_{\N}^{}
-2\!\int_{\S}^{\O}\!\!\!\d\ell\,\boldsymbol{\nabla}_{\!\!\perp}^{}\!\Phi_{\DE}^{}\,,~~~
\label{alpha}
\end{equation}
where $\,\boldsymbol{\nabla}_{\!\!\perp}^{}$ denotes the projection
of derivative $\,\boldsymbol{\nabla}\,$ onto the plane perpendicular
to the ray's path $\ell\,$.

Consider the matter part having mass $M$ with its mass density $\rho(\xx)$
distributed over a space region $\Omega(\xx_c^{})$,
where $\,\xx_c^{}\,$ is the position of its center of mass.
Thus, the total mass of matter is
$\,M=\int_{\Omega}^{}\!\d^3 \xx'\rho(\xx')$\,.\,
With this, we may express the Newtonian potential $\Phi_{\N}^{}$ as
\beqa
\label{eq:PhiN}
\Phi_{\N}^{}
=-\!\int_{\Omega}^{}\!\!\d^3 \xx'\frac{\rho(\xx')}{\,|\xx -\xx^{'}\!|\,}\,,
\eeqa
where the size of space $\Omega(\xx_c^{})$ is taken to be much smaller than
$\,|\xx -\xx_{c}^{}|$,\, and thus
$\,|\xx -\xx'|\simeq |\xx-\xx_c^{}|\,$.\,
Under the Newtonian potential $\Phi_{\N}^{}$, the deflection angle
$\vec{\alpha}_{\M}^{}$ of rays is determined by the matter contribution,
\beqa
\vec{\alpha}_{\M}^{}
= -2\!\int_{\S}^{\O}\!\!\!\d\ell\,
\boldsymbol{\nabla}_{\!\!\perp}^{}\!\Phi_{\N}^{}
\simeq -2M\!\!\int_{\S}^{\O}\!\!\!\d\ell\,
\frac{(\xx\!-\!\xx_{c}^{})_{\!\perp}^{}}{\,|\xx\!-\!\xx_{c}^{}|^{3}\,}\,.~~~~~
\label{alphadm}
\eeqa
The minus sign in front of Eq.(\ref{alphadm}) is notable, and
means that the deflecting direction is opposite to
$(\xx-\xx_{c}^{})_{\!\perp}^{}$.\,
Hence, the Newtonian potential tends to trap light rays.
This shows that the Newtonian gravitational potential
behaves as a convex lens.

\begin{figure}[t]
\centerline{
\includegraphics[width=14cm,height=6.5cm]{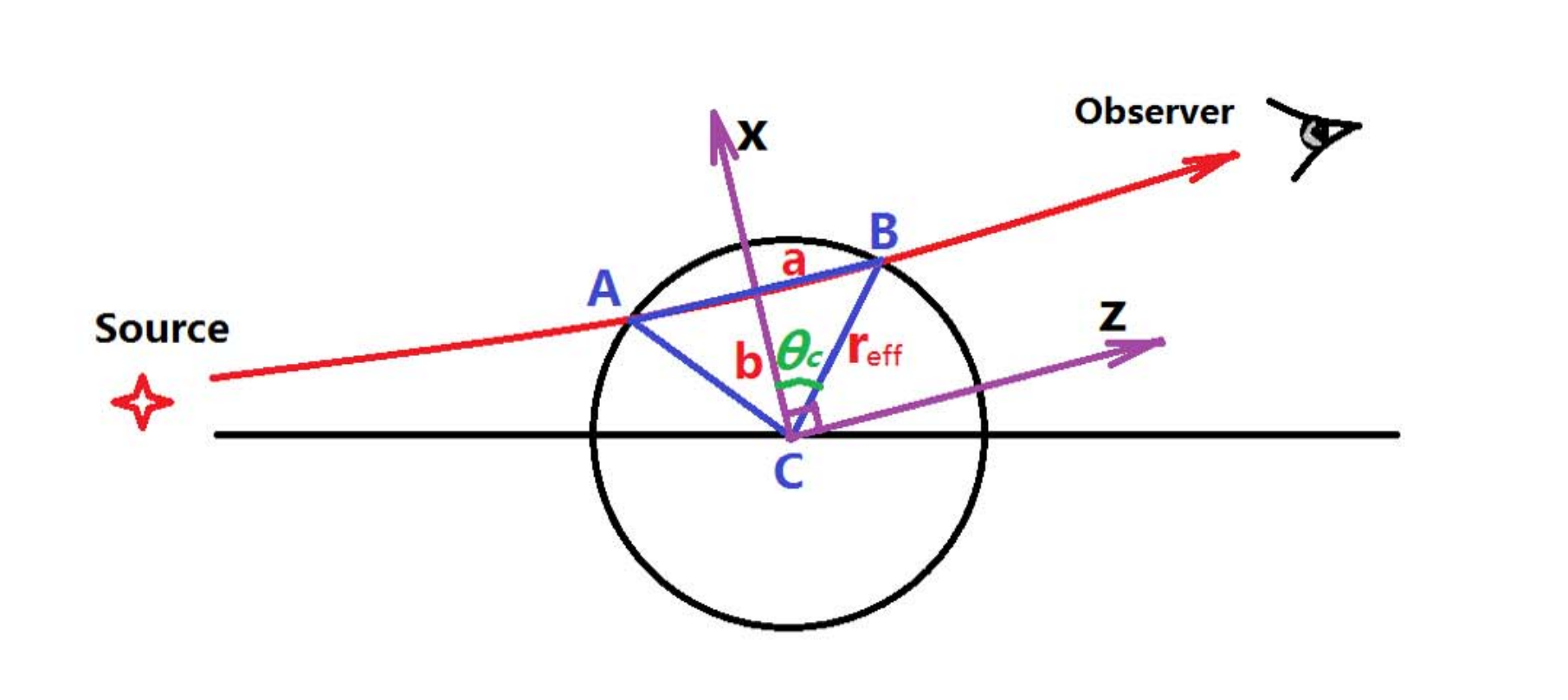}}
\vspace*{-5mm}
\caption{Concave gravitational lensing. The red curve represents the projected trajectory
of a ray in the $x$-$z$ plane, which is deflected by the gravitational potential
of dark energy. The circle region defines this lensing as an isolated gravitational system,
with radius $r_{\text{eff}}^{}$ and its center at the center of mass $(C)$.
The light ray path in the circle is well approximated
by a straight line segment $\over{AB}$, as the intersection between the path and the circle.
The shortest distance between $\overline{AB}$ and the center of mass $(C)$ is denoted
as $b$\,.\, The length $\,a=\frac{1}{2}|AB|$,\,
and the angle $\,\theta_c^{}=\frac{1}{2}\angle ACB\,$.}
\label{fig:DE}
\label{fig:2}
\end{figure}

Figure\,\ref{fig:2} illustrates a case where a ray passes through a region
within which the dark energy may dominate over the matter.
Here, the circle region defines this lensing as an isolated gravitational system
(within which the tidal effect from any other gravitational system can be ignored).
The circle has its center located at the center of mass of the matter $(C)$,\,
and has an effective radius $\,\reff\,$ which is
characetrized by the critical radius $\,\rc\,$
(at which the dark energy repulsive force just cancels the attractive Newtonian force,
cf.\ Sec.\,\ref{sec:4}).
The lensing parameter $\,b\,$ is the shortest distance
between the incident ray's path $\over{AB}$ and the point $C$.\,
For a given dark energy potential, denote its center by $\xx_{c}^{}$,\,
which coincides with the center of mass $(C)$.
From Eq.(\ref{potential}), we express the generic dark energy potential as
\beqa
\Phi_{\DE}^{}(r) \,=\,
-\!\(\!\frac{r^{2}_{\!o}}{\,|\xx\!-\!\xx_{c}|^{2}\,}\!\)^{\!\!\!\frac{3w +1}{2}}.
\label{eq:Phi-DE}
\eeqa
Thus, we have
$\,\boldsymbol{\nabla}_{\!\!\perp}^{}\!\Phi_{\DE}^{}
 = - (3w\!+\!1)
\frac{\,(\xx-\xx_c^{})_{\!\perp}^{}\,}{|\xx-\xx_c^{}|^{2}}\Phi_{\DE}^{}$.\,
Substituting this formula into Eq.(\ref{alpha}),
we derive the dark energy contribution
to the deflection angle,
\beqa
\Delta\vec\alpha_{\DE}^{}
&=& -2\!\int_{\S}^{\O}\!\!\!\d\ell\,
\boldsymbol{\nabla}_{\!\!\perp}^{}\!\Phi_{\DE}^{}
\nn\\
&=&
2(-3w\!-\!1)\,r^{3w+1}_{\!o}\!\!
\int_{\S}^{\O}\!\!\!\d\ell\,
\frac{(\xx\!-\!\xx_c^{})_{\!\perp}^{}}
     {\,|\xx\!-\!\xx_c^{}|^{3w+3}\,}.
\hspace*{9mm}
\label{alphade}
\label{eq:alphaDE}
\eeqa
This deflection angle $\,\Delta\vec\alpha_{\DE}^{}$\, is induced by
the dark energy potential \eqref{eq:Phi-DE} alone.
Note that the coefficient of Eq.\eqref{eq:alphaDE} is {\it positive}
due to $-(3w +1) > 0\,$,\, as required by the accelerated expansion
of the Universe \cite{BAO}.
This shows that dark energy deflects rays in the same direction as
$(\xx-\xx_c^{})_{\!\perp}^{}$.\,
Hence, the dark energy potential tends to diffuse
the rays and behaves as a concave lens.

\begin{figure}[t]
\centerline{
\includegraphics[width=14cm,height=6.5cm]{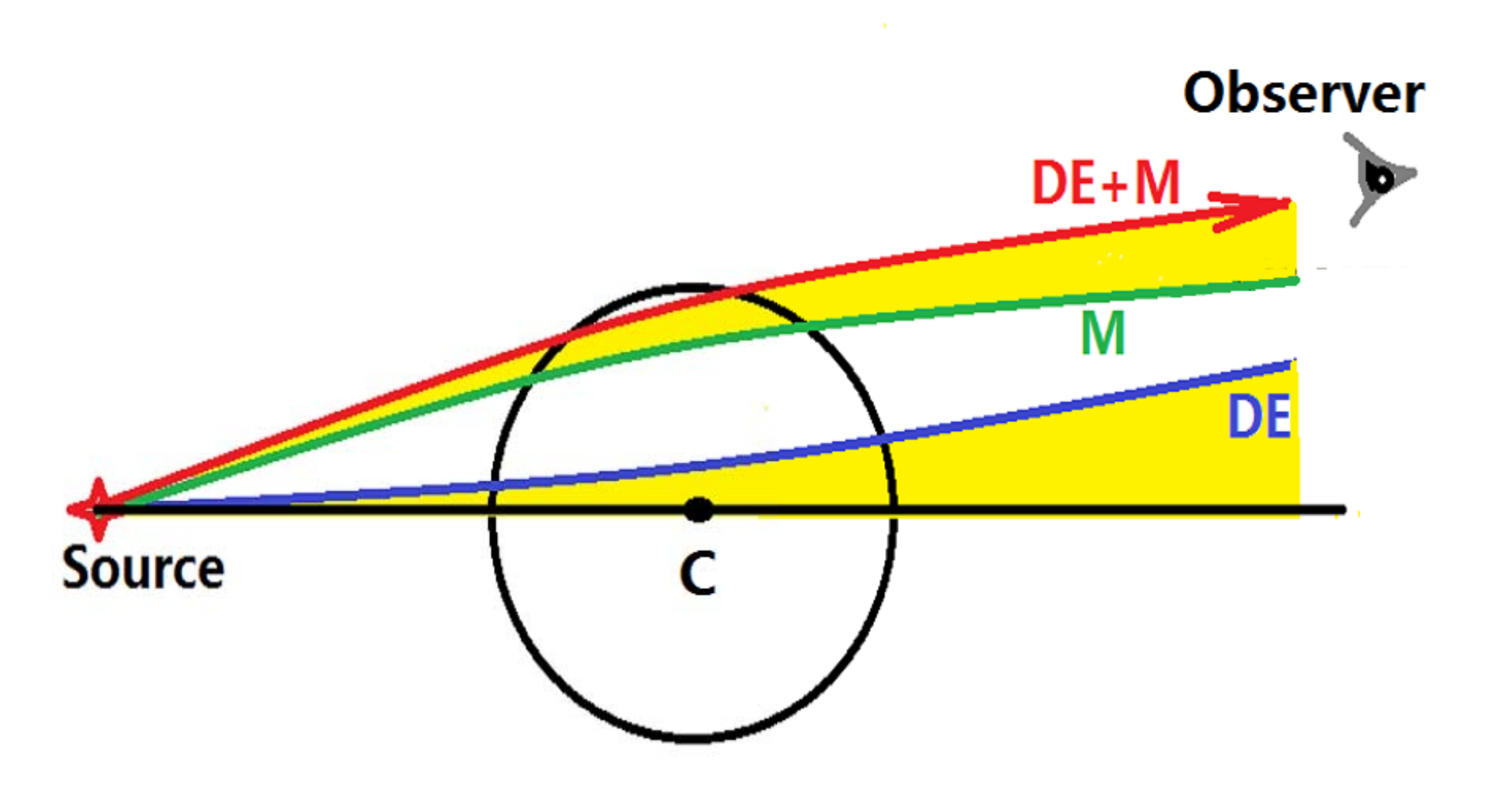} }
\vspace*{-3mm}
\caption{Gravitational lensing, including both contributions by the dark energy and matter.
The red curve represents the path of a ray. The green curve stands for the matter contribution
alone, and the blue curve illustrates the dark energy contribution alone.
The yellow region between the red and green curves shows the deflection effect
of the dark energy in the combined path.   }
\label{fig:DME}
\label{fig:3}
\end{figure}

In Fig.\,\ref{fig:3}, we illustrate that gravitational lensing
measures the combined effect of two forces:
the attractive force induced by matter and the repulsive force induced by dark energy.
The deflection angle is co-determined by these two forces
with opposite effects,
\beqa
\vec{\alpha}\,=\,
\vec{\alpha}_{\M}^{}+\Delta\vec{\alpha}_{\DE}^{} \,.
\label{alphat}
\label{eq:alpha=M+DE}
\eeqa
For the projections onto any given $x$ axis, we have
\beqa
\text{sign}(\Delta\alpha^{x}_{\DE})
\,=\, -\text{sign}(\alpha^{x}_{\M}),
\label{eq:alpha-sign}
\eeqa
which means that the rays are deflected
by matter and dark energy in the opposite directions.
The dark energy always acts as a concave lens and diffuses the rays,
while the matter behaves as a convex lens and attracts the rays.
Fig.\,\ref{fig:3} shows that the Newtonian potential of matter acts as a convex lens
and deflects the incident ray along the green curve.
On the other hand, the dark energy potential
will diffuse the ray along the blue curve. The combined effect is to deflect
the ray along the path of the red curve.
The net effect induced by dark energy is the shift region
(marked in yellow color) between the red curve and the green curve.
Measuring this effect can provide a direct evidence for dark energy.

The deflection angle in Eq.\eqref{eq:alphaDE} takes a {\it general form,} where
the equation of state parameter $\,w\,$ plays an important role
and can describe different models of dark energy when it takes their corresponding
values (cf.\ Appendix-\ref{app}).
Hence, making precision measurement on the deflection effect \eqref{eq:alphaDE}
via gravitational lensing can directly probe the existence of dark energy, and will
further discriminate the equation of state parameter $\,w\,$ among
different dark energy models.

\vspace*{1mm}

We can follow the standard procedure to write down the lens equation.
The angular diameter distances
from the source ($\S$) to the observer ($\O$),
from the source ($\S$) to the lens ($C$), and
from the lens ($C$) to the observer ($\O$) are denoted as
$D_{\S\O}^{}$, $D_{{\S} C}^{}$, and $D_{C\O}^{}$
in the FLRW spacetime, respectively.
Considering the large scales
$\,D_{\S\O}^{},D_{{\S} C}^{},D_{C\O}^{}\gg \rc$\,
and the small-angle relations, we have the lens equation,
\beqa
\vec{\theta}\,\simeq\,
\vec{\beta}+\vec{\alpha}\,\frac{D_{{\S} C}^{}}{\,D_{\S\O}^{}\,}
\,=\, \vec{\beta}+\!\(\vec{\alpha}_{\M}^{}\!+\!\Delta\vec{\alpha}_{\DE}^{}\)
\!\frac{D_{{\S} C}^{}}{\,D_{\S\O}^{}\,} \,,
\label{eq:LensEq}
\eeqa
where the angle $\vec{\beta}$ corresponds to the position of the unlensed source
$\S$, the angle $\vec{\theta}$ corresponds to the apparent position,
and the deflection angle
$\,\vec{\alpha}=\vec{\alpha}_{\M}^{}+\Delta\vec{\alpha}_{\DE}^{}$\,
is given by Eq.\eqref{eq:alpha=M+DE}.
As we will show in Sec.\,\ref{sec:4}-\ref{sec:5A}
as well as Appendices\,\ref{app:A4} and \ref{app:B2},
this deflection angle $\,\vec{\alpha}\,$ remains unchanged for the regions
outside the deflection region.
The form of the lens equation \eqref{eq:LensEq} agrees to
Refs.\,\cite{Bartelmann99}\cite{Ishak2007d}.
It is clear that in Eq.\eqref{eq:LensEq}, the matter and dark energy contributions
to the deflection angle
$\,\vec{\alpha}\,(=\vec{\alpha}_{\M}^{}+\Delta\vec{\alpha}_{\DE}^{})$\,
are {\it on equal footing.}  Hence, the dark energy deflection $\vec{\alpha}_{\DE}^{}$
cannot be attributed to the angular diameter distances separately
while leaving the matter contribution $\vec{\alpha}_{\M}^{}$ unaffected.
This point was also mentioned in \cite{Ishak2010} for the case of $w=-1$.

\vspace*{1mm}

Before concluding this section, we comment on the special limit $\,M\!\to 0$\,.\,
In this limit, the lensing setup is removed, while the dark energy remains
everywhere in the Universe. So one may wonder whether the
dark energy potential $\,\Phi_{\DE}^{}\,$ could still deflect
any incident ray passing through the space. The answer is no,
and the explanation is instructive.
Inspecting the gravitational potential $\,\PhiN \!+ \PhiDE\,$ in
Eq.\eqref{eq:Phi} and the spacetime metric \eqref{eq:dS2} or \eqref{eq:dSS2},
we see that the origin point of the coordinate frame for this isolated lensing system
is chosen to be the center of mass point $C$ of $M$
(cf.\ Figs.\,\ref{fig:2}-\ref{fig:3}),
and the potential $\,\PhiN \!+ \PhiDE\,$
is consistently solved from the Einstein equation
\eqref{eq:EQ} as shown in Appendix\,\ref{app:0}.
But, in the limit $\,M\!\to 0\,$,\, the current center of mass point $C$ no longer
has its meaning. Hence, we have to redefine the isolated system by including
the source of this incident ray, where the source must have a
nonzero mass, say $M'$,\, whose center of mass point may be denoted as $C'$.\,
Now, for the consistency of solving Einstein equation,
we must redefine a new coordinate frame with its origin at the point $C'$,
and then rederive the potential $\,\PhiN +\PhiDE$\,  in this frame
as Appendix\,\ref{app:0}.
The key point is to note: because the incident ray is emitted from its
source $M'$ centered at $C'$,\, it propagates along the {\it radial direction} of
the new potential form $\,\PhiN \!+ \PhiDE\,$.\, Hence, no deflection on this ray
could be generated according to the definition \eqref{alpha},
and we can directly infer the deflection angle from
Eqs.\eqref{alphadm} and \eqref{eq:alphaDE},
$\,\vec{\alpha}_{\M}^{}=\Delta\vec{\alpha}_{\DE}^{}=0\,$.\,
This shows that the current formulation is fully consistent.

\vspace*{3mm}
\section{\hspace*{-1.5mm}Probing Concave Lensing of Dark Energy}
\label{sec:del}
\label{sec:4}
\vspace*{1mm}

When a ray passes through the deflection region, the gravitational field
can bend the ray by a small deflection angle.
The ray's path $\ell$ over the deflection region may be approximated
by a straight line segment $\over{AB}$\,,\, as in Fig.\,\ref{fig:2}.
For convenience, we choose a new coordinate system in Fig.\,\ref{fig:2},
where the $z$-axis is parallel to the straight line $\over{AB}$\,.\,
Here we only show the lensing effect for dark energy.
In this coordinate system, the point on the line segment $\over{AB}$\,
is denoted as $\,z=z(\ell )$. From Eq.(\ref{alphade}), we have
\beqa
\Delta\vec\alpha_{\DE}^{} \,=\,
2(-3w\!-\!1)\,r^{3w+1}_{\!o}\!
(\xx\!-\!\xx_c^{})_{\!\perp}^{}\!\!\int_{A}^{B}\!\!\!
\frac{\hspace*{-6mm}\d z}
     {\,|\xx\!-\!\xx_c^{}|^{3w+3}\,}.
\hspace*{6mm}
\label{alphade2}
\label{eq:alphaDE-AB}
\eeqa

To simplify the analysis, we assume that the deflector is an isolated and
spherical system, in which all the matter can be approximated as
a spherically symmetric mass-distribution with total mass $M$.\,
The gravitational potential is described by Eq.(\ref{potential}).
The recent combination of BAO, SN and CMB data gives\,\cite{BAO},
$\,w=-0.97\pm 0.05\,$ for $w$CDM (constant $w$ with a flat universe),
which is fairly close to $\,w=-1\,$ and leads to
$\,3w+1<0\,$ at $12.7\sigma$ level and
$\,3w+2<0\,$ at $6.1\sigma$ level.
Thus, Eq.(\ref{potential}) shows that the dark energy force is always repulsive
[cf.\ Eq.\eqref{eq:force}]
and increases with the distance $\,r\,$.\,
When $\,r\,$ reaches a critical value $\,r_{\text{cri}}^{}\,$,\,
this repulsive force will balance the attractive Newtonian force, where
the net force acting on matter vanishes. Hence, beyond $\,\rc\,$ the matter
will begin to escape the gravitational bonds.
From the gravitational potential (\ref{potential}),
we deduce both the Newtonian force
and dark energy force as in Eq.\eqref{eq:force} of Appendix\,\ref{app:11}.
Hence, we can derive the critical radius $\,\rc\,$
(for a generical state parameter $\,w\,$),\,
\beqa
\rc \,=\, \ro\(\!\frac{M}{\,|3w\!+\!1|\,\ro\,}\!\)^{\!\!-\frac{1}{3w}} ,
\label{rc}
\label{eq:rc}
\eeqa
at which the dark energy force balances the Newtonian attraction.
For the special case of $\,w=-1$\,,\, Eq.\,\eqref{eq:rc} reduces to
$\,\rc =({3M}/{\Lambda})^{1/3}$,\, which coincides with \cite{Ho15}.

In the region $\,r\ll\rc\,$,\, the gravitational force is dominated
by the Newtonian attraction of matter. In the region with $\,r>\rc$\,,\,
the repulsive dark energy force will dominate over the Newtonian attraction.
The post-Newtonian approximation holds well for
$\,r\lesssim\rc\,$.\,
The weak field approximation requires the effective radius $\,\reff \ll \ro$\,.\,
Also, the size of $\,\reff\,$ is properly chosen to avoid tidal effect
from any other gravitational object, so that the lensing system
behaves as an isolated system.
To have nonnegligible dark energy effect on the lensing,
we consider $\,\reff\,$ around the same order of $\,\rc\,$,\, i.e.,
the ratio $\,\neff \equiv \reff /\rc = O(1)\,$.\,
In the region with $\,\rc\lesssim r \lesssim \reff\,$,\, one may still make estimates
under the post-Newtonian approximation.

The spacetime region $\,r\lesssim\reff\,$ of this isolated astrophysical system
can be smoothly embedded into a nearly flat FLRW universe
($r>\reff$) \cite{Darmois1927}\cite{ES1945}.
Thus, we require that the matter density of such an isolated lensing system
equals the averaged matter density $\,\rho_M^{}\,$
at cosmological scales from the CMB measurements.
This will also ensure the Einstein equation
to hold across the boundary ($\,r=\reff\,$) \cite{Darmois1927}\cite{ES1945}.
Hence, we have the matching condition
\beqa
\label{eq:matching}
\frac{M}{~\frac{4\pi}{3}\rrreff~}
\,=\,
\rho_M^{} \,,
\eeqa
where we have\,\cite{Planck2015},
$\,\rho_M^{}=\rho_{\DE}^{}\(\Omega_{\M}^{}/\Omega_{\DE}^{}\)\simeq 0.44 \rho_{\DE}^{}\,$.\,
From Eqs.\eqref{eq:matching} and \eqref{eq:rc}, we derive
\beqa
\label{eq:reff-DE}
\reff \,=\, \(\!\frac{3M}{\,4\pi\rho_M^{}\,}\!\)^{\!\!\frac{1}{3}}
=\, \rc\!\(\!\frac{\,3|3w\!+\!1|\,}{\,4\pi r^2_o\,\rho_M^{}\,}\!\)^{\!\!\frac{1}{3}}
\!\!\(\!\frac{\,\rc\,}{\ro}\!\)^{\!\!|w|-1}.
\eeqa
We may give an explicit estimate on the size of
the effective radius $\,\reff\,$.\,
For the case of the cosmological constant as dark energy ($w=-1$),
we derive the dark energy density $\,\rho_{\DE}^{}=\rho_{\Lambda}^{}\,$
as follows,
\beqa
\label{eq:rho-CC}
\rho_{\Lambda}^{} \,=\, \frac{\Lambda}{\,8\pi\,}
\,=\, \frac{\,3M\,}{~8\pi\,\rccc\,} \,.
\eeqa
Thus,  using Eqs.\eqref{eq:reff-DE}-\eqref{eq:rho-CC} with $\,w=-1$\,,\,
we can estimate the size of $\,\reff\,$
in terms of the critical radius $\,\rc\,$,
\beqa
\label{eq:reff--rc}
\reff \,=\, \rc\!
\(\!\frac{2}{\,\Omega_{\M}^{}/\Omega_{\Lambda}^{}\,}\!\)^{\!\!\frac{1}{3}}
\,\simeq\, 1.7\,\rc \,,
\eeqa
and $\,\neff = \reff /\rc \simeq 1.7 =O(1)$,\,
as expected. This is a fairly reasonable estimate.
The effective radius $\,\reff\,$ may serve
to characterize the transition scale between the
Schwarzschild-de\,Sitter (SdS)
spacetime inside the deflection region (\,$r\lesssim \reff$\,)
and the FLRW spacetime beyond the lensing system (\,$r> \reff$\,).
We will further discuss the embedding of the isolated lensing system into
the flat FLRW spacetime in Appendices\,\ref{app:A4} and \ref{app:B2}.

According to Fig.\,\ref{fig:2}, we can express the lensing parameters
$(a,\,b)$ as,
$\,a=\reff\,\sin\theta_c^{}$\, and
$\,b=\left|(\xx-\xxc )_{\!\perp}^{}\right|
    =\reff\,\cos\theta_c^{}$\,.\,
Thus, from Eq.\eqref{eq:alphaDE-AB},
we compute the size of the deflection angle produced by dark energy,
$\,\Delta\alpha_{\DE}^{}=|{\Delta\vec\alpha_{\DE}^{}}|\,$,
\beqa
\Delta\alpha_{\DE}^{}
& = & 2|3w\!+\!1|\,r_{\!o}^{3w+1}b\!\int_{-a}^{+a}\!\!
\frac{\d z~~~~~~~~~~}{\,(b^2\!+\!z^2)^{(3w+3)/2}\,}
\nn\\[2mm]
& = & 4|3w\!+\!1|\!
\(\!\frac{r_{\!o}^{}}{\reff}\!\)^{\!\!3w+1}\!\!\!\!
\int_{0}^{\theta_c^{}}\!\!\!\d\theta
\(\!\frac{\cos\theta}{\cos\theta}_c^{}\!\)^{\!\!3w+1}\!,
\hspace*{12mm}
\label{alphavalue}
\label{eq:alphaDE-w}
\eeqa
where $\,\theta_c^{}=\arctan(a/b)< \frac{\pi}{2}$\,.\,
Along the same path, from Eq.(\ref{alphadm}),
we compute the size of matter-induced deflection angle
$\,\alpha_{\M}^{}=|\vec\alpha_{\M}^{}|\,$ as follows,
\beqa
\alpha_{\M}^{}\,=\,
\frac{\,4M\,}{\,\reff\,}\!\int_{0}^{\theta_c^{}}\!\!\!
\d\theta\,\frac{\cos\theta}{\,\cos\theta_c^{}\,}
\,=\,\frac{\,4M\,}{\,\reff\,}\tan\theta_c^{}\,.
\hspace*{4mm}
\label{alphadm2}
\eeqa
For the special case of fixing $\,b \,(=\reff \cos\theta_c^{})$\,
and taking $\,\reff\!\to\!\infty\,$ (with $\theta_c^{}\!\to\!\frac{\pi}{2}$),
we see that the formula \eqref{alphadm2} recovers the conventional result
of light bending\,\cite{Einstein}\cite{Weinberg72},
$\,\alpha_{\M}^{}=\frac{\,4M\,}{\,b\,}\,$.\,
Hence, the size of the total deflection angle is
$\,\alpha =|\vec{\alpha}|=
 |\vec{\alpha}_{\M}^{}\!+\!\Delta\vec{\alpha}_{\DE}^{}|\,$.\,
From Eq.\eqref{eq:alpha-sign}, we see that the dark energy contribution
always tends to cancel the matter contribution to the deflection angle
$\,\alpha\,$, and will produce a deficit as compared to the conventional
lensing prediction without dark energy.

As an explicit demonstration, let us consider the cosmological constant model
of dark energy. It corresponds to $\,w=-1\,$ and
$\,\ro =\sqrt{{6}/{\Lambda}}\,$ in the gravitational potential \eqref{eq:Phi}.
From Eq.\eqref{eq:rc}, the lensing system in this case has a critical radius
$\,\rc =({3M}/{\Lambda})^{1/3}$\,.\,
For the region within the effective radius
$\,r\lesssim\reff\!\sim\!\rc\,$,\,
the post-Newtonian approximation holds well.
With Eq.\eqref{eq:alphaDE-w}, we compute the dark energy contribution
to the deflection angle,
\beqa
\label{eq:alphaDE-CC}
\Delta\alpha_{\DE}^{}
\,=\, 4\(\!\!\frac{\,\reff\,}{\ro}\!\!\)^{\!\!2}\!
\sin 2\theta_c^{} \,=\,
2n_{\text{eff}}^2\sin 2\theta_c^{}\!
\(\!\!\frac{\,\Lambda M^2}{3}\!\)^{\!\!\frac{1}{3}},
\eeqa
where $\,\neff = \reff /\rc =O(1)$.\,
The maximal value
$\,\Delta\alpha^{\max}_{\DE} =4({\reff}/{\ro})^2
   =2n_{\text{eff}}^2(\Lambda M^2\!/3)^{\!{1}\!/{3}}$\,
is achieved at
$\,\theta_c^{}=\,\frac{\pi}{4}$\,.\,

We further clarify the special limit of $\,M\!\to 0\,$.\,
From the expression $\,\rc =({3M}/{\Lambda})^{1/3}$,\, we find that the limit
$\,M\!\to 0\,$ leads to $\,\reff \sim \rc \propto M^{\frac{1}{3}}\!\to 0\,$.\,
Thus, Eq.\eqref{eq:alphaDE-CC} gives
$\,\Delta\alpha_{\DE}^{}\propto \rff \propto M^{\frac{2}{3}}\!\to 0\,$
for each given angle $\,\theta_c^{}\,$.
For a general state parameter $w$\,,\,
we can deduce from Eqs.\eqref{eq:alphaDE-w} and \eqref{eq:rc},
$\,\Delta\alpha_{\DE}^{}\propto r_{\text{eff}}^{-(3w+1)}
\propto r_{\text{cri}}^{-(3w+1)}
\propto M^{(1+\frac{1}{3w})}\,$,\,
in which $\,1+\frac{1}{3w}>0\,$ holds at $12.7\sigma$ level as required
by the existing observational data\,\cite{BAO}. Hence, the zero mass limit
$\,M\!\to 0\,$ always enforces $\,\Delta\alpha_{\DE}^{}\to 0\,$.\,
This also agrees to our conclusion at the end of Sec.\,\ref{sec:3},
and shows that our approach is fully consistent.

For the current case of single lensing and
in the regions beyond the isolated lensing system with
$\,\rrr \gg \rfff\!\sim\!\rccc\,$,\,
the Newtonian potential becomes negligible, and
we explicitly show in Appendices\,\ref{app:A4} and \ref{app:B2}
that the spacetime geometry conformally recovers the FLRW metric
with vanishing spatial curvature ($K=0$),
which will not change the deflection angle
of the passing light ray. Hence, the total lensing effect of dark energy
is given by Eq.\,\eqref{eq:alphaDE-CC}.

From Eqs.\eqref{alphadm2} and \eqref{eq:alphaDE-CC}, we further derive the ratio,
\beqa
\frac{\,\Delta\alpha_{\DE}^{}\,}{\alpha_{\M}^{}}
\,=\, n^3_{\text{eff}}\cos^2\!\theta_c^{}
=\, \neff \frac{b^2}{\,r^2_{\text{cri}}\,},
\hspace*{4mm}
\label{alphadeodm}
\eeqa
where $\,\neff = \reff/\rc =O(1)\,$.\,
This means that when the lensing parameters $(b,\,\reff )$ are comparable
to the critical radius $\,\rc\,$,\, the dark energy correction
$\,\Delta\alpha_{\DE}^{}\,$ will be significant as compared to
the matter contribution $\,\alpha_{\M}^{}\,$.\,
Hence, for the case of $\,b,\reff\sim\rc\,$,\,
measuring gravitational lensing effect
can directly probe the dark energy via its produced deficit in the deflection angle
$\,\alpha =|\vec{\alpha}_{\M}^{}\!+\!\Delta\vec{\alpha}_{\DE}^{}|\,$.\,

The cosmological constant model of dark energy corresponds to $\,w=-1\,$ and
$\,\ro =\sqrt{6/\Lambda}\,$ in Eq.\,\eqref{eq:Phi}.
Thus, we can estimate $\,\ro\simeq 0.72\times 10^{4}\,$Mpc.\,
For an astrophysical object, such as galaxy or galaxy cluster,
its typical mass is about
$10^{12-16}M_{\odot}^{}$,\, where $M_{\odot}^{}$ is the solar mass.
We estimate its critical radius
$\,\rc =({3M}/{\Lambda})^{1/3}\simeq (1.1-23)$\,Mpc.\,
The  effective radius $\,\reff =\neff\,\rc\,$,\,
where the coefficient $\,\neff =O(1)\,$ depends on the definition of the lens
as an isolated system.
For an ideal isolated lens,
our matching estimate \eqref{eq:reff--rc} gives $\,\neff\simeq 1.7\,$.\,
Thus, from Eq.\eqref{eq:alphaDE-CC},
we estimate the dark energy contribution,
$\Delta\alpha_{\DE}^{\max}\simeq
 n_{\text{eff}}^2\!\times\!( 0.018''\!-8.5'')$.\,
This is already a significant effect, as we may recall that the
conventional prediction of the matter-induced light bending through the sun is
$\,\alpha_{\odot}^{}=1.75''$ \cite{Einstein}\cite{Weinberg72}.

\begin{center}
\begin{figure}[t]
\hspace*{-5mm}
\includegraphics[width=8.5cm,height=6.6cm]{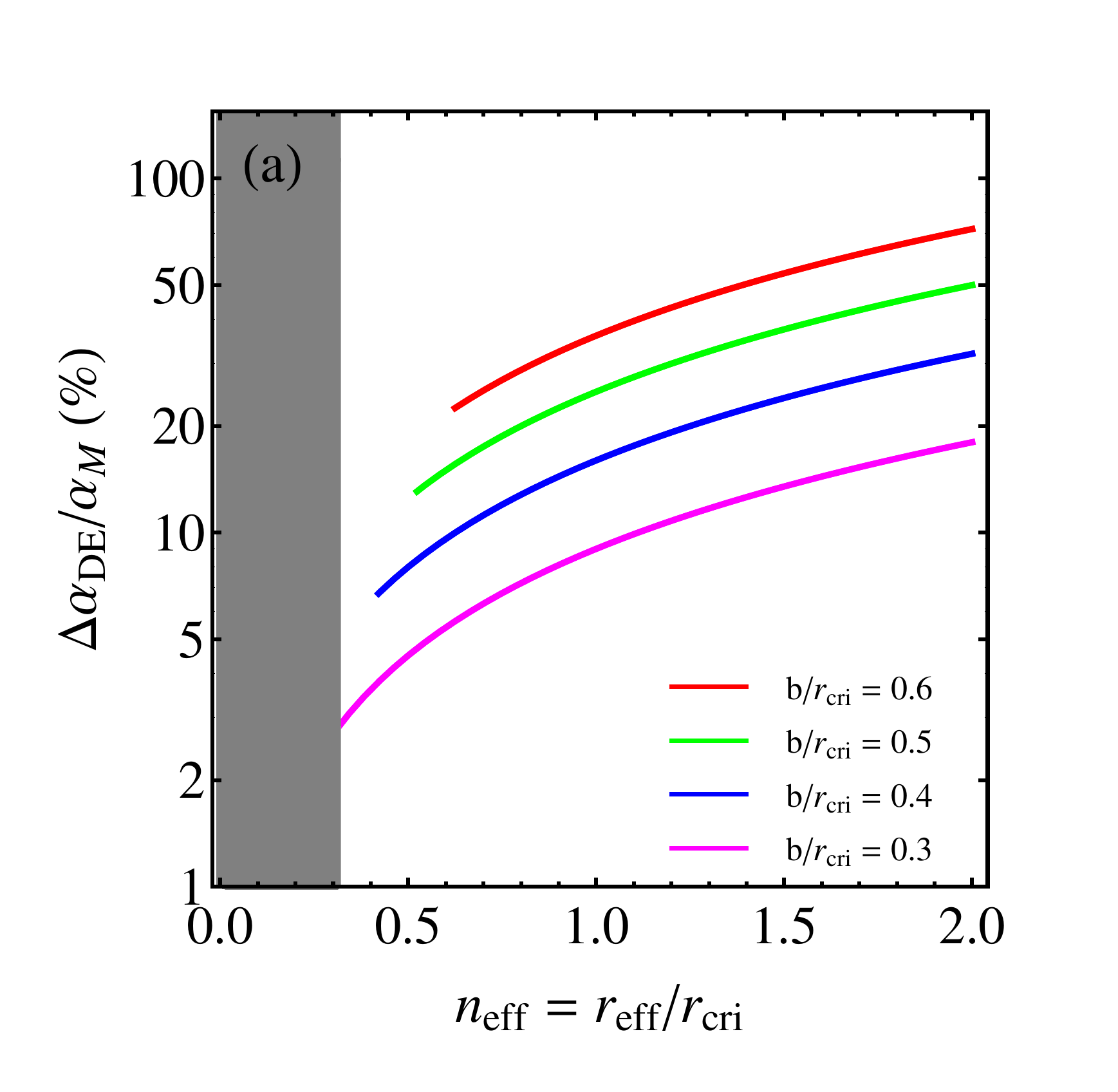}
\hspace*{-10mm}
\includegraphics[width=8.5cm,height=6.6cm]{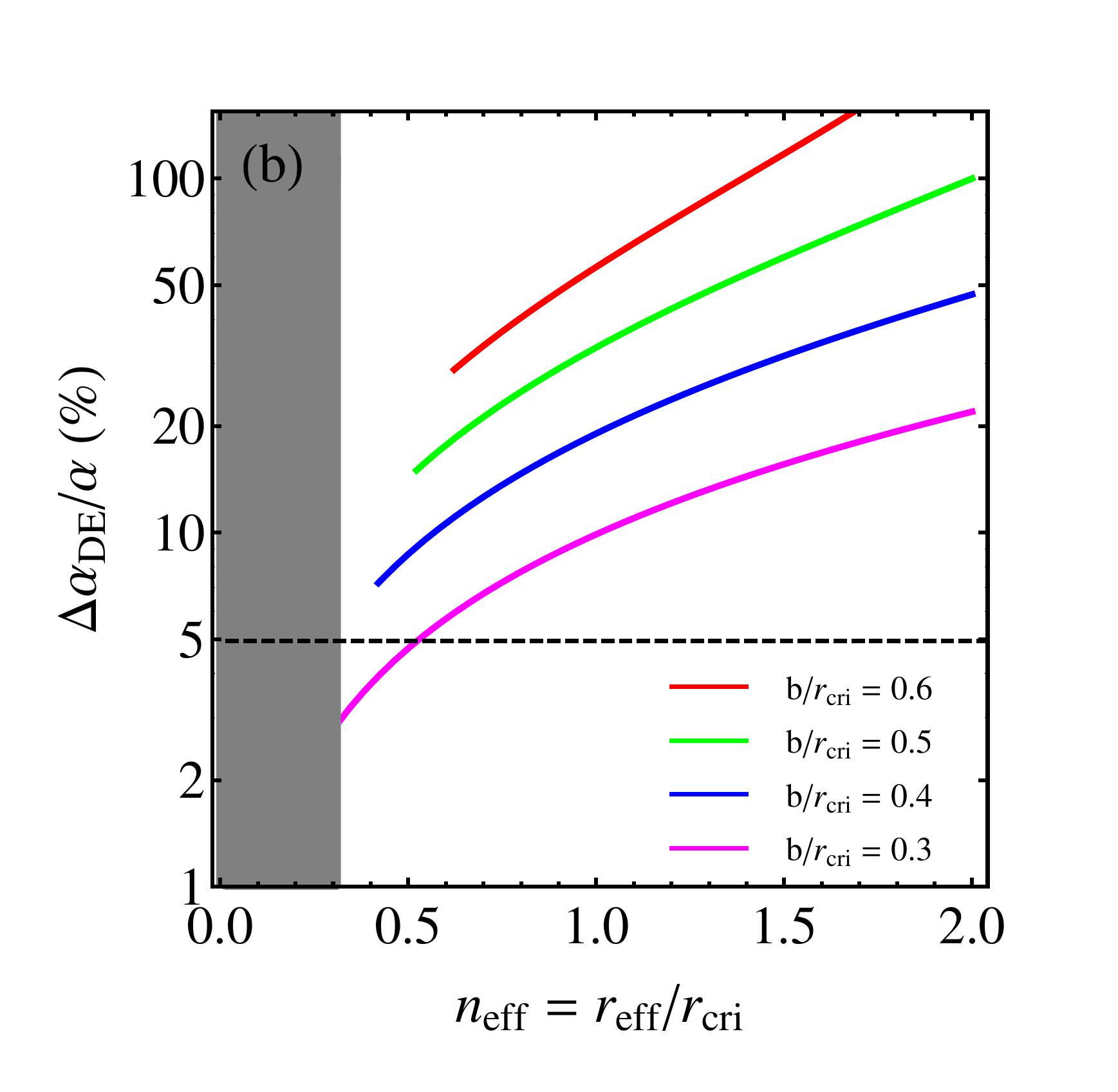}
\hspace*{-3mm}
\vspace*{-4mm}
\caption{Dark energy contribution $\,\Delta\alpha_{\DE}^{}$ to the deflection angle
$\,\alpha\,$ of the gravitational lensing. 
Plot\,(a) shows $\,\Delta\alpha_{\DE}^{}$ relative to the matter contribution
$\Delta\alpha_{\M}^{}$,\, while plot\,(b) depicts the ratio
$\,\Delta\alpha_{\DE}^{}/\alpha$\,.\,
An accuracy of 5\% on the $\,\alpha\,$ measurement can be reached
by the current lensing experiments\,\cite{Zitrin11,Sluse12,Umetsu16},
as illustrated by the horizontal dashed line in plot\,(b).}
\label{fig:Obs1}
\label{fig:4}
\end{figure}
\end{center}

The current lensing experiments for galaxies at redshift $\,z=0.1-0.5$\,
\cite{Malin2014}\cite{Shu2016}
can measure the deflection angle to a relative accuracy of
$\,\Delta\alpha/\alpha\sim 5\%$,\,
and measure $\,{\Delta}\alpha\sim 0.01''$.\,
For galaxies or galaxy clusters at redshift $\,z=0.5-2$\,
\cite{Zitrin11,Sluse12,Umetsu16}, the current measurements
can reach $\,\Delta\alpha/\alpha\sim 5\%$\,
and $\,{\Delta}\alpha\sim 0.05''$.\,
For the actual experimental measurements, a 5$\%$ accuracy of $\,\alpha\,$
is expected to be on the conservative side, as we will choose in Fig.\,\ref{fig:4}
for comparison.

In passing, we also note that Ishak {\it et al} \cite{Ishak2007d} independently considered
some typical lensing systems
with galaxy cluster masses $\,M=(1.37-13.8)\!\times\!10^{13}M_{\odot}^{}$,\,
and used a different method (formula)
to estimate the cosmological constant contribution
to the light bending angle,
$\,\Delta\alpha_{\Lambda}^{}\simeq (0.005''\!-0.025'')$.
For the conventional matter-induced light bending angle $\alpha_{\M}^{}$
at the leading order,
Ref.\cite{Ishak2007d} agrees to our Eq.\eqref{alphadm2}
with its distance $R$ (between the centers of the deflection region and the lens)
related to our lensing parameters via $\,R=\reff/\tan\theta_c^{}$\,.\,
For the above mass-range of typical lenses, they chose
the deflection region with
$\,R\simeq (0.06-0.14)\text{Mpc}\simeq (0.02-0.03)\rc\ll \reff\sim \rc$.\,
Since we choose the deflection region with
$\,r\leqq \reff = O(\rc )\,$ and
our Eq.\eqref{eq:alphaDE-CC} gives
$\,\Delta\alpha_{\DE}^{}\propto r_{\text{eff}}^2\sim r_{\text{cri}}^2\,$,
it is expected that for $\,\reff = \rc (\,\gg\! R)\,$,\,
we obtain a much larger bending angle
$\,\Delta\alpha_{\DE}^{\max}\simeq (0.11''\!-0.49'')$
for the same lensing mass-range of \cite{Ishak2007d}.
If we choose the effective radius $\reff$ around its lower limit
$\,\reff \simeq 0.3\,\rc\,$ (cf.\ the discussion below),
this means a smaller deflection region
and we deduce a smaller bending angle accordingly,
$\,\Delta\alpha_{\DE}^{\max}\simeq (0.009''\!-0.044'')$.

\vspace*{-12mm}

In Fig.\,\ref{fig:4}, we illustrate the dark energy contribution
$\,\Delta\alpha_{\DE}^{}\,$ to the deflection angle $\,\alpha\,$
of the gravitational lensing.
To be concrete, we realize the dark energy via cosmological constant.
Plot\,(a) presents the ratio of $\,\Delta\alpha_{\DE}^{}\,$ over the matter
contribution $\,\alpha_{\M}^{}\,$,\, while plot\,(b) depicts the relative ratio
$\,\Delta\alpha_{\DE}^{}/\alpha\,$.\,
Using Eqs.\eqref{eq:alphaDE-CC} and \eqref{alphadeodm},
we vary the ratio $\,\neff = \reff/\rc\,$ for the horizontal axis,
and choose $\,b/\rc =(0.3,\, 0.4,\, 0.5,\, 0.6)$\, as the benchmark points,
corresponding to the curves from bottom to top in each plot.
For the shaded region of $\,\reff/\rc < 0.3$\,,\, the matter
sector could not well behave as an isolated gravitational system,
so it will be excluded from the analysis.
In the deflection region of
$\,0.3\,\rc \lesssim r \lesssim \reff$\,,
there is almost no matter distribution and the Newtonian potential
of matter can be well approximated as generated by a point-like object.
For instance, the Milky Way has its dark matter halo extended out to the
galactocentric distance of $\sim\! 200$\,kpc \cite{Bhattacharjee14},
which is about $\,0.2\,\rc\,$ for $\,\rc \!\approx 10^3\,$kpc.\,
So it is safe to choose a larger distance $\,\sim 0.3\,\rc\,$
as the lower limit of effective radius $\,\reff\,$.\,
In the deflection region of
$\,0.3\,\rc \lesssim \,r \lesssim \,\reff = O(\rc)$,\,
the effect of dark energy contribution may be extracted
by subtracting the background of matter contribution.
This subtraction does not depend on the detail of matter distribution since
it behaves point-like.
Taking an accuracy of 5$\%$ for the present lensing measurements,
we see that in the region of $\,\neff > 0.3$,\,
the predicted curves in plot\,(b) are mostly
above the dashed black horizontal line
(denoting the 5$\%$ accuracy), and are thus testable.

We note that the dark energy lensing effect \eqref{eq:alphaDE-w}
is fully expressed in terms of the dark energy parameters
$(w,\,\ro )$ besides the lensing parameters $\,\reff\,$ and
$\,\theta_c^{}=\arccos(b/\reff)\,$.\,
Hence, Eq.\eqref{eq:alphaDE-w} holds for both the cosmological constant
model ($w=-1$) and other dark energy models, such as the quitessence
model ($-1<w<-\frac{1}{3}$) and the phantom model ($w<-1$).\,
Thus, making precision lensing measurements is important for probing the
dark energy contribution $\,\Delta\alpha_{\DE}^{}\,$ to the deflection angle,
and for further discriminating dark energy models with different values
of the state parameter $w$\,.\,
Eq.\eqref{eq:alphaDE-w} shows that
for a given gravitational lensing system, the dark energy correction
$\,\Delta\alpha_{\DE}^{}\,$ is fully determined by the two generic parameters
$(w,\,\ro )$.\, So it is possible to fit $(w,\,\ro )$ from analyzing
the variation of luminance or distortion of the 2d image of background light sources
due to the concave lensing of dark energy. With extracted information about
the state parameter $\,w$\,,\, we could identify the proper model of dark energy.
The current telescopes appear sensitive to probing this parameter
via gravitational lensing.

\vspace*{3mm}
\section{\hspace*{-1.5mm}Light Orbital Equation with Generic State Parameter $\mathbf{w}$}
\label{sec:5A}
\label{sec:55}
\vspace*{1mm}

In this section, considering a gravitational lensing system of total mass $M$
(with a point-like spherical mass-distribution),
we first derive the light orbital equation for dark energy models with
a generic state parameter $\,w$\, in the SdS$w$ spacetime.
For the special case of the cosmological constant
model of dark energy ($w=-1$),\, it reduces to the light orbital equation
in the familiar SdS spacetime.
At the end of this section,
we also clarify the difference of our approach from the literature
concerning the light bending effect induced by $\,\Lambda\,$.

From the metric \eqref{eq:dSS2} in Appendix\,\ref{app:11} and
using the null geodesic condition $\,\d S^2 = 0$\,,\,
we derive the following light orbital equation
with a generic state parameter $\,w$\,,
\beqa
\label{eq:lorbit1}
\left(\frac{\d u}{\d\phi}\right)^{\!\!2}
=~\frac{1}{\,b^2\,}-u^2+2Mu^3+2\,{\ro}^{\!\!3w+1}u^{\!3(w+1)} \,,
\hspace*{9mm}
\eeqa
for $\,\theta\,=\frac{\pi}{2}$\,.\, In the above,
$\,u={1}/{r}$,\, $\,b=L/E$\,,\, $b$ is the impact parameter,
and $E$ and $L$ are total energy and angular momentum, respectively.
We will give the derivation of Eq.\eqref{eq:lorbit1} in Appendix\,\ref{app:B1}.
The cosmological constant model of dark energy predicts $\,w=-1$\,,\,
and in this case our lensing system corresponds to the SdS spacetime.
We note that for $\,w=-1$,\,
the last term on the right-hand-side of Eq.\eqref{eq:lorbit1}
becomes a pure constant, independent of $\,r$\,.\,
Thus, we can reexpress Eq.\eqref{eq:lorbit1} as follows,
\beqa
\label{eq:1stOE-B}  
\left(\frac{\d u}{\d\phi}\right)^{\!\!2}
=~\frac{1}{\,b^2\,}+\frac{2}{\,r_o^2\,} -u^2 + 2Mu^3
+ \frac{2}{\,r_o^2\,}\!\left[(\ro u)^{3(w+1)}\!-1 \right] ,
\hspace*{9mm}
\eeqa
where the last term on the right-hand-side vanishes for
the cosmological constant dark energy ($w=-1$),\, and
gives nonzero $r$-dependent contribution for other dark energy models
with $\,w\neq -1$.\,

From Eq.\eqref{eq:lorbit1}, we can derive
a second-order ordinary differential equation (ODE),
\beqa
\label{eq:lorbit2}
\frac{{\d}^2 u}{\d{\phi}^2}
\,=\, -u+3Mu^2+3(w\!+\!1)\,{\ro}^{\!\!3w+1}u^{\!3w+2} \,,
\hspace*{9mm}
\eeqa
which holds for a general state parameter $\,w\,$.\,
Under the weak deflection assumption, the orbital equation
can be approximated by introducing a small perturbation
to the undeflected straight line in the flat space.
In Eq.\eqref{eq:lorbit2}, the $w$-dependent terms arise
from the correction of dark energy, which vanishes for $\,w=-1$\,.\,
It shows that only for the special case of the cosmological constant ($w=-1$),\,
Eq.\eqref{eq:lorbit2} reduces to the following form,
\beqa
\label{eq:lorbit2CC}
\frac{{\d}^2 u}{\d{\phi}^2} \,=-u+3Mu^2\,,
\hspace*{9mm}
\eeqa
which agrees to \cite{Ishak2007} and happens to take the same form as the
conventional light orbital equation with $\,\Lambda = 0$\,.\,

But, the original orbital equation \eqref{eq:lorbit1}
(which is the first-order ODE)
does depend on the dark energy term even for the case of $\,w=-1\,$,
\beqa
\label{eq:lorbit1CC}
\left(\frac{\d u}{\d\phi}\right)^{\!\!2} \,=\,
\frac{1}{\,b^2\,}-u^2+2Mu^3+\frac{\,\Lambda\,}{3} \,.
\eeqa
It may be rewritten as\,\cite{Arakida12},
\beqa
\label{eq:lorbit1CC2}
\(\frac{\d u}{\d\phi}\)^{\!\!2}
\,=\, \frac{1}{\,B^2\,}-u^2+2Mu^3\,,\,
\eeqa
with an effective impact parameter $\,B\,$ defined via
$\,\frac{1}{\,B^2\,}=\frac{1}{\,b^2\,}+\frac{\,\Lambda\,}{3}\,$.\,
But, for all other cases with $\,w\neq -1\,$,\,
the $w$-dependent dark energy term
in Eq.\eqref{eq:lorbit1} or Eq.\eqref{eq:1stOE-B} does depend
on the coordinate $\,r\,$,\, and thus {\it cannot} be absorbed into the
redefinition of an effective impact parameter $B$.\,
This means that making precision measurement on the light deflection
can discriminate between dark energy models with $\,w= -1\,$ and
$\,w\neq -1\,$.\,
Furthermore,
Eq.\eqref{eq:lorbit1CC2} holds only for the point-like light source,
rather than the more realistic 2-dimensional (2d) light source.
For such 2d light source, the impact parameter $b$ varies its value
for different positions on the surface of the source.
Hence, no universal effective impact parameter $B$ exists
to fully absorb $\Lambda$ effect in such realistic cases.
Thus, fitting the lensing data from 2d light sources can
discriminate the deflection effect induced by $\Lambda$\,.

\vspace*{4mm}
\section{\hspace*{-1.5mm}Conclusions}
\label{sec:5}
\label{sec:66}
\vspace*{1.5mm}

The mystery of dark energy poses a great challenge to modern science.
Various on-going and future experiments\,\cite{DEexp} are making enormous efforts
to probe the origin and nature of dark energy.
In this work, we showed that gravitational lensing
can serve as an important tool for direct probe of dark energy at astrophysical scales.
For an isolated astrophysical system (such as the galaxy or galaxy cluster),
we derived a general form for the repulsive potential
of dark energy with a generic equation of state parameter $\,w\,$,\,
and computed its direct lensing effect under the post-Newtonian approximation.
We demonstrated that the dark energy acts as a {\it concave lens,} contrary to
the convex lensing effect of both visible and dark matter.
Hence, measuring this concave lensing effect can directly probe the existence
and nature of dark energy.

In section\,\ref{sec:2}, under the post-Newtonian approximation,
we presented the repulsive power-law potential \eqref{eq:Phi} for dark energy,
containing a generic equation of state parameter $\,w\,$ which can
describe various dark energy models.
This is derived in the Schwarzschild-de\,Sitter spacetime
with general $w$\, (denoted as SdS$w$ metric).
Then, in section\,\ref{sec:3}, with the potential \eqref{eq:Phi},
we derived the dark energy contribution
$\,\Delta\vec{\alpha}_{\DE}^{}\,$ to the deflection angle
$\,\vec{\alpha}\,$ of a gravitational lensing system in Eq.\eqref{eq:alphaDE}.
We further proved in Eq.\eqref{eq:alpha-sign} that this deflection always
takes opposite sign to the matter contribution $\,\vec{\alpha}_{\M}^{}\,$
along any given direction.
Hence, the dark energy effect always acts as a concave lens, in contrast with
the conventional matter effect, as shown in Fig.\,\ref{fig:2} and Fig.\,\ref{fig:3}.

In section\,\ref{sec:4}, for a general state parameter $\,w\,$,\,
we derived the critical radius $\,\rc\,$ in Eq.\eqref{eq:rc},
at which the dark energy force balances the Newtonian attraction.
For a given gravitational lensing with mass $M$,
we imposed the matching condition \eqref{eq:matching}
and deduced its effective radius $\,\reff\,$
in Eqs.\eqref{eq:reff-DE} and \eqref{eq:reff--rc},
which defines this lens as an ideal isolated system.
Eq.\eqref{eq:reff--rc} explicitly shows $\,\reff =O(\rc)\,$.\,
Then, we computed the size of the dark-energy-induced deflection angle,
$\,\Delta{\alpha}_{\DE}^{}\,$,\, as given by Eq.\eqref{eq:alphaDE-w}.
We demonstrated that the dark-energy-induced deflection angle
$\,\Delta\alpha_{\DE}^{}\propto M^{(1+\frac{1}{3w})}\,$
(with $1\!+\!\frac{1}{3w}>0$\,), which increases with the lensing mass $M$ and
consistently vanishes in the zero-mass limit $\,M\!\!\to\!0$\,.\,
We explicitly derived the prediction of $\,\Delta{\alpha}_{\DE}^{}\,$
in Eq.\eqref{eq:alphaDE-CC} for the cosmological constant model of dark energy
($w\!=\!-1$).\,
We computed the ratio between the dark energy and matter contributions to the
deflection angle in Eq.\eqref{alphadeodm}. In Fig.\,\ref{fig:4}(a)-(b),
we presented the relative ratios $\,\Delta{\alpha}_{\DE}^{}/\alpha_{\M}^{}\,$
and $\,\Delta{\alpha}_{\DE}^{}/\alpha\,$,\, as functions of the effective radius
$\,\reff\,$ (in unit of $\,\rc\,$) and for different benchmarks of the lensing parameter
$\,b$\,.\, We estimated this dark energy lensing effect and found that
the current gravitational lensing experiments are already sensitive to
the direct probe of dark energy.
In section\,\ref{sec:5A}, for an isolated lensing system,
we presented a new orbital equation \eqref{eq:lorbit1} [or \eqref{eq:1stOE-B}] of light rays
including the generic state parameter $w$\,,\,
and discussed its implication for the observations. 
We gave a systmetical derivation of Eq.\eqref{eq:lorbit1} in Appendix\,\ref{app:B}. 
We showed in Appendix\,\ref{app:C1} that  
the cosmic expansion effect is negligible
within a lensing system at typical astrophysical scales.
For regions outside the lensing system, we proved 
the approximate conformal flatness of the SdS metric and the SdS$w$ metric
for $\,r^3/\rccc\gg 1\,$,\, as in Appendix\,\ref{app:44} and Appendix\,\ref{app:C2}. 
So there is no leading correction to the deflection angle studied in Sec.\,\ref{sec:3}-\ref{sec:4}.

Finally, we conclude that in general it is important to make
the precision measurements on the dark-energy-induced deflection effect \eqref{eq:alphaDE-w}
via astrophysical gravitational lensing, which could directly probe the nature of
dark energy and further discriminate the equation of state parameter $\,w\,$
among different dark energy models.

\vspace*{5mm}
\noindent
{\bf\large Acknowledgements}
\\[1mm]
We thank Xingang Chen, Stephen Hsu, Mustapha Ishak,
Yipeng Jing, Mauro Sereno, Yu-Chen Wang, and Jun Zhang for useful discussions.
We also thank Xuelei Chen, Zuhui Fan and Pengjie Zhang for related discussions.
This research was supported in part by the National NSF of China
(under Grants 11275101, 11135003, 11675086).



\vspace*{3mm}
\begin{appendix}
\section{\hspace*{-1.5mm}Dark Energy Potential and Dark Energy Models}	
\label{app}
\label{app:A}


In Appendix\,\ref{app:11},
we first use Einstein equation to derive the generic dark energy
potential \eqref{eq:Phi} for an isolated astrophysical system.
Then, we discuss its connection with the typical dark energy models\,\cite{DEmodel},
in Appendix\,\ref{app:22}-\ref{app:44},
including the phantom model, the quintessence model, and the cosmological constant model.
In Appendix\,\ref{app:44}, we will further show that
in the regions beyond the isolated lensing system with $\,\RRR \gg 1\,$,\,
the Newtonian potential becomes negligible, and
the spacetime geometry reduces to the de\,Sitter metric
which conformally recovers the flat FLRW metric
with vanishing spatial curvature ($K=0$).
So the region $\,\RRR \gg 1\,$
will not change the deflection angle of the passing light ray.

\vspace*{2mm}
\subsection{\hspace*{-1.5mm}General Gravitational Potential Including Dark Energy}	
\label{app:0}
\label{app:11}
\label{app:A1}
\vspace*{1mm}

The static and spherically symmetric spacetime can be described
by the following form,
\beqa
\d S^2 =\, e^{\mathbb{A}}\d t^2\! -e^{\mathbb{B}}\d r^2\!
- r^2\!\(\d \theta^2\!+\sin^2\!\theta\,\d\phi^2\)\!,
\hspace*{10mm}
\label{eq:dSS}
\eeqa
where $\,\mathbb{A} =\mathbb{A}(r)\,$ and $\,\mathbb{B} =\mathbb{B}(r)$\,
are radial functions.
For conventions, we choose the Minkowski metric 
$\eta_{\mu\nu}^{}=\eta^{\mu\nu}=\text{diag}(1,-1,-1,-1)$,\,
and set the constants $\,G =h=c=1$\, throughout as mentioned above Eq.\eqref{eq:Phi}.
Thus, the Einstein equation\,\cite{Einstein} takes the form
\beqa
\label{eq:EQ}
R_{\mu\nu}^{}-\frac{1}{2}g_{\mu\nu}^{}R
\,=\, -8\pi\, T_{\mu\nu}^{}\,,
\eeqa
where the energy-momentum tensor $\,T_{\mu\nu}^{}\,$
contains contributions from both matter and dark energy.
We can use the metric \eqref{eq:dSS} to express the Einstein equation
\eqref{eq:EQ} as follows,
\begin{eqnarray}
\hspace*{-2mm}
& 8\pi\,{T_t}^t& \!= - e^{-\mathbb{B}}\!\left(\frac{1}{r^2} -
\frac{\mathbb{B}^\prime}{r}\right)\! + \frac{1}{r^2}\,,
\\[1mm]
\hspace*{-2mm}
& 8\pi\,{T_r}^r& \!= - e^{-\mathbb{B}}\!\left(\frac{1}{r^2}
+ \frac{\mathbb{A}^\prime}{r}\right)\! + \frac{1}{r^2} \,,
\hspace*{10mm}
\\[1mm]
\hspace*{-2mm}
& 8\pi\,{T_\theta}^\theta& \!= 8\pi\,{T_\phi}^\phi
= - \frac{\,e^{-\mathbb{B}}}{2}
\!\(\!\mathbb{A}^{\prime\prime}\!+\!
\frac{{\,\mathbb{A}^\prime}^2}{2} \!+\!
\frac{\,\mathbb{A}^\prime\!\!-\!\mathbb{B}^\prime\,}{r} \!-\!
\frac{\,\mathbb{A}'\mathbb{B}'\,}{2}\!\)\!,
\hspace*{14mm}
\end{eqnarray}
where \,$\mathbb{A}' = {\rm d}\mathbb{A}(r)/{\d}r$\, and
\,$\mathbb{B}' = {\d}\mathbb{B}(r)/{\d}r$\,.

The condition of linearity and additivity requires $\,{T_t}^t = {T_r}^r\,$ \cite{Kiselov03},
which leads to $\,\mathbb{A} + \mathbb{B} = \text{constant}$.
Here, one can fix the gauge of $\,\mathbb{A} + \mathbb{B} = 0$\,
for the static coordinate system without losing generality,
since the constant factor can be absorbed by a proper rescaling of time.
Setting $\,\mathbb{B} = -\ln(1\!+\! F)$,\, we have the following,
\begin{eqnarray}
4\pi{T_t}^t  &=&  4\pi{T_r}^r = - \frac{1}{\,2r^2\,}\, (F \!+ rF^\prime)\,,
\label{At}\\[1mm]
4\pi{T_\theta}^\theta  &=&  4\pi{T_\phi}^\phi = - \frac{1}{\,4r\,}\,(2 F^\prime
\!+rF^{\prime\prime})\,.
\hspace*{10mm}
\label{Ath}
\end{eqnarray}
The condition of linearity and additivity means that
the linear sum of the possible solutions of $F$ is still the solution
of Eqs.\eqref{At}-\eqref{Ath}.

For the dark energy models,
the equation of state parameter $\,w\,$
is a constant and $\,w < -\frac{1}{3}$ \,\cite{BAO}.\,
The pressure $\,p\,$,\, the energy density $\,\rho\,$,\,
and the state parameter $\,w\,$ are connected by
\begin{eqnarray}
\label{ADE1}
w \,=\,\frac{\,p\,}{\,\rho\,} \,.
\end{eqnarray}
We can express the energy-momentum tensor for dark energy
as follows\,\cite{Kiselov03},
\begin{eqnarray}
{T_t}^t &=& \rho(r) \,,
\label{AparaB1}
\\[1mm]
\label{AparaB}
{T_i}^j  &=&  3\,w\, \rho(r)\!
\left[-(1\!+\!3\,\B)\frac{r_i^{}\,r^j}{\,r_n^{}r^n\,}+\B\,{\delta_i}^j\right]\!,
~~~~~~~~
\end{eqnarray}
where $\,\B\,$ will be determined by the dark energy spacetime.
We see that the space-component is proportional to the time-component.
Averaging over the angles gives,
\beqa
\label{eq:<Tij>}
\left\langle {T_i}^j\right\rangle = - \rho(r)\,w\,
{\delta_i^{}}^j = - p(r)\,{\delta_i^{}}^j\,,~~~~
\eeqa
where \,$\langle r_i^{}r^j\rangle = \frac{1}{3}\, {\delta_i^{}}^j r_n^{} r^n$
\cite{Kiselov03}.
The above formulas could include matter contribution
with $\,\rho_{\M}^{}=M\delta^3(\vec{r})$,\,
which has no contribution to $(p,\,w)$ for $\,r\neq 0\,$.\,

Using the condition of linearity and additivity can fix the parameter in
Eq.\eqref{AparaB},
%
$\,\B = -(3 w\!+\! 1)/(6w)$,\,
which leads to  \cite{Kiselov03}
\begin{eqnarray}
\label{eq:T1-DE}
{T_t}^t  &=&  {T_r}^r = \rho \,,
\label{Atq}
\\[1mm]
\label{eq:T2-DE}
{T_\theta}^\theta  &=&  {T_\phi}^\phi
= -\frac{\,\rho\,}{2}\,(3 w+1)
= -\frac{1}{2}(3p+\rho )\,.
\label{Athq}
\end{eqnarray}

Combining Eqs.(\ref{At})-(\ref{Ath}) with the expressions
(\ref{Atq})-(\ref{Athq}) leads to the equation,
\beqa
r^2F'' + 3 (w\!+\!1)\,r\,F' + (3 w\!+\!1)F   \,=\, 0\,,
\eeqa
which has the following solutions,
\begin{eqnarray}
F_{\DE}^{}  =  -2\left(\! \frac{\,\ro\,}{r}\!\right)^{\!\!3w+1} \!,
\hspace*{5mm}
F_{\rm N}^{}  =  -2\frac{\,r_g^{}\,}{r},~~~~~
\label{eq:FF}
\end{eqnarray}
where $\,\ro\,$ and $r_g^{}$ are the integral constants serving as the
normalization scale-factors.
Thus, the sum of the above solutions
$\,F=F_{\rm N}^{}+F_{\DE}^{}\equiv 2\Phi\,$ is also the solution
due to the linearity and additivity. Hence, we can express the metric parameters
$\,\mathbb{A}=-\mathbb{B}=\ln (1\!+\!F) = \ln(1\!+\!2\Phi)\,$ as follows,
\beqa
\label{eq:AB}
e^{\mathbb{A}} &=& e^{-\mathbb{B}}
\,=\, 1+2\Phi\,,
\\[1.5mm]
\label{eq:PhiNDE}
\Phi &=&
-\frac{\,r_{g}\,}{r} - \left(\! \frac{\,\ro\,}{r}\!\right)^{\!\!3w+1}
\equiv\, \Phi_{\N}^{}\! + \Phi_{\DE}^{}\,,
\hspace*{13mm}
\eeqa
where in Eq.\eqref{eq:PhiNDE} the first term $\,\Phi_{\N}^{}\,$
just corresponds to the Newtonian potential
with mass $\,r_g^{}=M$,\, and the constant
$\,\ro$\, in the $\,\Phi_{\DE}^{}$\, term
characterizes the size of the present universe.

Substituting Eqs.\eqref{eq:AB}-\eqref{eq:PhiNDE} into Eq.\eqref{eq:dSS},
we derive the spacetime metric,
%
\beqa
\label{eq:dSS2}
\dis
\d S^2 \,=\, (1\!+\!2\Phi)\d t^{2}
\!-\!\frac{\d r^2}{\,(1\!+\!2\Phi)\,}\!
-\! r^2\!\(\d \theta^2\!+\sin^2\!\theta\,\d\phi^2\)\!,
\hspace*{10mm}
\eeqa
with the gravitational potential $\,\Phi =\Phi_{\N}^{}\! + \Phi_{\DE}^{}\,$,
\beqa
\label{eq:Phi-NDE2}
\Phi_{\N}^{}  =  -\frac{\,M\,}{r}\,, ~~~~~
\Phi_{\DE}^{} =  -\!\left(\! \frac{\,\ro\,}{r} \!\right)^{\!3w+1} \!.
\hspace*{9mm}
\eeqa
The above spacetime metric \eqref{eq:dSS2} reproduces
Eq.\eqref{eq:metric1} up to linear order of $\Phi$
under the post-Newtonian approximation,
and the potential \eqref{eq:Phi-NDE2} coincides with Eq.\eqref{eq:Phi}.
For the special case of cosmological constant ($w=-1$) as dark energy,\,
the metric \eqref{eq:dSS2} reduces to that of the familiar
Schwarzschild-de\,Sitter (SdS) spacetime.
For convenience, we will denote Eqs.\eqref{eq:dSS2}-\eqref{eq:Phi-NDE2}
as SdS$w$ metric.

The general formulas \eqref{eq:dSS2}-\eqref{eq:Phi-NDE2} give
the following Newtonian gravitational force and dark energy force,
\beqa
\label{eq:force}
\vec{F} \,=\, -\boldsymbol{\nabla}\Phi \,=
\(\!-\frac{\,M\,}{\,r^2\,}
+\frac{|3w\!+\!1|}{\ro}\!\!\(\!\frac{r}{\ro}\!\)^{\!\!|3w\!+\!2|}\)\!
\boldsymbol{\hat{r}}\,,
\hspace*{10mm}
\eeqa
where the current data\,\cite{BAO} requires
$\,3w+1 < 0\,$ at $\,12.7\sigma$\, level and
$\,3w+2 < 0\,$ at $\,6.1\sigma$\, level.
Hence, the dark energy force is always repulsive
and increases with distance $\,r\,$.\,
Requiring the cancellation between the Newtonian and dark energy forces
$\,\vec{F}=0\,$,\,
we derive the critical radius,
\beqa
\rc \,=\, \ro\(\!\frac{M}{\,|3w\!+\!1|\,\ro\,}\!\)^{\!\!-\frac{1}{3w}}.
\label{Arc}
\label{eq:Arc}
\eeqa
This reproduces Eq.\eqref{eq:rc} in Sec.\,\ref{sec:4}.
At the scale of $\,r \lesssim \rc\,$,\,
the matter contribution is larger than that of dark energy,
and the matter distribution has significant structures.
Thus, at such scales, the cosmological principle no longer holds
and the FLRW metric is not a valid description of the spacetime.
This means that for $\,r \lesssim \rc\,$,\,
it is not justified to use FLRW metric with matter density
treated as small perturbation. As we demonstrated above,
for the scale $\,r \lesssim \rc\,$,\, the spacetime geometry is best described
by the SdS$w$ metric \eqref{eq:dSS2} [or \eqref{eq:metric1}] with the
gravitational potential \eqref{eq:Phi-NDE2} [or \eqref{eq:Phi}].
On the other hand, with the key concept of critical radius
\eqref{eq:Arc}, we should define the (large) cosmological scale as
$\,\rrr\gg\rccc\,$,\, at which the cosmological principle will be realized.
With the astronomical data, we may estimate the maximal critical radius
$\,r_{\text{cri}}^{\max}\,$ for the galaxy cluster with the largest mass
$\,M_{\max}^{}$,\, which sets the lower limit on the cosmological scales.

For the matter part, we have the conventional Poisson equation,
\beqa
-\nabla^2\Phi_{\M}^{} =\, 4\pi(\rho_{\M}^{}\!+3\,p_{\M}^{}),~~~~
\label{eq:M00}
\eeqa
where the pressure $p_{\M}^{}$ is negligible compared to the energy density
$\rho_{\M}^{}$.

With Eqs.\eqref{eq:T1-DE}-\eqref{eq:T2-DE},
the energy-momentum tensor of dark energy takes the following form,
\beqa
\left\langle {T_{\mu}^{}}^\nu\right\rangle
=\, \text{diag}(\rho_{w}^{},-p_{w}^{},-p_{w}^{},-p_{w}^{}),~~~~
\label{eq:TE00}
\eeqa
where $\,\rho_{w}^{}=\rho_{\DE}^{}$ is the dark energy density and $\,p_{w}^{}=p_{\DE}^{}$
denotes its pressure, with the state parameter $w$\,.
From Eq.\eqref{At} and Eqs.\,\eqref{AparaB1},\eqref{eq:Phi-NDE2},
we derive
\beqa
- \nabla^2\Phi_{\DE}^{} =\,
-4\pi\rho_w^{}\,.~~~~
\label{eq:DE00a}
\eeqa
Thus, we may combine Eqs.\eqref{eq:M00} and \eqref{eq:DE00a},
\beqa
-\nabla^2\Phi =\, 4\pi(\rho_{\M}^{}\!-\rho_w^{}) \,,
\hspace*{10mm}
\label{eq:DE00}
\eeqa
where we have dropped the pressure term $\,p_{\M}^{}\,$ since
$\,p_{\M}^{}\ll\rho_{\M}^{}$\, for non-relativistic Newtonian source $M$.\,
Eq.\eqref{eq:DE00} is the modified Poisson equation, which includes
the contributions from both Newtonian gravity and dark energy.
For the case of cosmological constant as dark energy, we have
$\,w=-1\,$ and $\,\rho_w^{}=\frac{\Lambda}{\,4\pi\,}\,$.\,
Thus, in this special case,
the Poisson equation \eqref{eq:DE00} reduces to the form,
\beqa
-\nabla^2\Phi =\, 4\pi\rho_{\M}^{}\!-\Lambda \,,
\hspace*{10mm}
\label{eq:CC00}
\eeqa
which agrees with \cite{Ho15}.

In general, the potential $\,\Phi_{\DE}^{}$\, in Eq.\eqref{eq:Phi-NDE2}
can describe different dark energy models characterized by
the equation of state parameter:\,
$-1<w <-\frac{1}{3}\,$ for the quintessence model,
$\,w<-1$\, for the phantom model, and
$\,w=-1\,$ for the cosmological constant model.


\vspace*{2mm}
\subsection{\hspace*{-1.5mm}Phantom Model}	
\label{app:1}
\label{app:22}
\label{app:A2}
\vspace*{1mm}

In the phantom model, the equation of state parameter
$w_p^{}$ is a constant with $\,w_p^{} < -1$\,.\,
The pressure $\,p_p^{}\,$,\, the energy density $\,\rho_p^{}\,$,\,
and the state parameter $\,w_p^{}\,$ are given by
\begin{eqnarray}
\label{pdp1}
\rho_p^{} &=& -\frac{\,\dot{\phi}^2}{2}+V(\phi),
\\[1mm]
\label{pdp2}
p_p^{} &=& -\frac{\,\dot{\phi}^2}{2}-V(\phi),
\\[1mm]
\label{pdp3}
w_p^{} &=& \frac{\,p_p^{}\,}{\,\rho_p^{}\,},
\end{eqnarray}
where $\,V(\phi)\,$ is the potential of phantom field.
From Eq.\eqref{eq:Phi-NDE2}, we have the dark energy potential
\begin{eqnarray}
\Phi_{p}^{}  =  -\left( \frac{\,\ro\,}{r} \!\right)^{\!3w_p+1} .
\end{eqnarray}

\vspace*{2mm}
\subsection{\hspace*{-1.5mm}Quintessence Model}	
\label{app:2}
\label{app:33}
\label{app:A3}
\vspace*{1mm}

In the quintessence model,
the pressure $\,p_q^{}\,$,\, the energy density $\,\rho_q^{}\,$,\,
and the state parameter $\,w_q^{}\,$ are expressed as follows,
\begin{eqnarray}
\rho_q^{} &=& \frac{\,\dot{\phi}^2}{2} +V(\phi) \,,
\\[1mm]
p_q^{} &=& \frac{\,\dot{\phi}^2}{2} -V(\phi) \,,
\\[1mm]
w_q^{} &=& \frac{\,p_q^{}\,}{\rho_q^{}} \,,
\end{eqnarray}
where $\,V(\phi)\,$ is the potential of quintessence field.

After exchanging $\rho_q^{}$ with $-p_p^{}$,\,
and $p_q^{}$ with $-\rho_q^{}$,\,
and following the above derivation, we have
\begin{eqnarray}
\rho_q^{} &\leftrightarrow & \,-p_p^{}\,, \hspace*{6mm}
p_q^{}\leftrightarrow\,-\rho_p^{}\,,
\\[2mm]
w_q^{} \!&=& \!\frac{\,p_q^{}\,}{\rho_q^{}}
\,=\, \frac{\,-\rho_p^{}\,}{-p_p^{}} \,=\, \frac{1}{\,w_p^{}} \,.
\end{eqnarray}
For the quintessence spacetime, the state parameter
$\,w_q^{}\,$ is a constant within the range
$\,-1< w_q^{} < -\frac{1}{3}\,$.

Similar to the case of phantom model, we have the quintessence dark energy potential
from Eq.\eqref{eq:Phi-NDE2},
\begin{eqnarray}
\Phi_q^{}  =  -\left( \frac{\,\ro\,}{r} \right)^{\!3w_q^{}+1} .
\end{eqnarray}

\vspace*{2mm}
\subsection{\hspace*{-1.5mm}Cosmological Constant Model}	
\label{app:3}
\label{app:44}
\label{app:A4}
\vspace*{1mm}

In the case of the cosmological constant model of dark energy ($\Lambda>0$),
the equation of state parameter is given by $\,w=-1\,$.\,
Thus, from the general form in Eq.\eqref{eq:Phi-NDE2},
we have the dark energy potential $\,\Phi_{\DE}^{}\,$ reduced to
\begin{equation}
\label{eq:Phi-CC}
\Phi_{\Lambda}^{}=-\(\!\frac{r}{\,\ro\,}\!\)^{\!\!2}
=-\frac{\,\Lambda\,}{6}r^{2} \,,
\end{equation}
with the scale factor $\,\ro =\sqrt{{6}/{\Lambda}}$\,,\,
which is consistent with \cite{Ishak2010}\cite{Ho15}.

For the cosmological constant model of $\Lambda>0$\,,\,
the generic spacetime metric \eqref{eq:dSS2} reduces to
the SdS geometry,\footnote{%
Eq.\eqref{eq:dSS2A4} is called the Schwarzschild-anti-de\,Sitter metric
for $\Lambda<0$,\, and is known as Kottler metric\,\cite{Kottler}
for a generic cosmological constant $\Lambda$\, which can be either positive
or negative.}
\beqa
\label{eq:dSS2A4}
\dis
\d S^2 = \(\!1\!-\!\frac{2M}{r}\!-\!\frac{\,\Lambda r^2\,}{3}\!\)\!\d t^{2}
-\frac{\d r^2}{\,(1\!-\!\frac{2M}{r}\!-\!\frac{\,\Lambda r^2\,}{3}\!)\,}
- r^2\!\(\d \theta^2\!+\sin^2\!\theta\,\d\phi^2\)\!.
\hspace*{10mm}
\eeqa
In Sec.\,\ref{sec:4}, we explicitly derived in Eq.\eqref{eq:alphaDE-CC}
the cosmological constant contribution $\,\Delta\alpha_{\DE}^{}\,$ to the light deflection
in the region $\,r\lesssim\reff\!\sim\!\rc\,$ of the isolated lensing system.
In the following, we analyze the region beyond this lensing system
with $\,\RRR \gg 1\,$,\,
which corresponds to $\,\Phi_{\N}^{}/\Phi_{\DE}^{}\sim \rccc / \rrr \ll 1\,$.\,
For this region, the Newtonian potential term $\Phi_{\N}^{}$
becomes negligible, and thus the SdS geometry \eqref{eq:dSS2A4} reduces to the
de\,Sitter metric,
\beq
\label{eq:dSS2A4c}
\ba{lcl}
\d S^2 &\simeq\,& \dis \d S_{\text{dS}}^2
\,=\, \(\!1\!-\!\frac{\,\Lambda r^2}{3}\!\) \!\d t^{2}
\!- \frac{\d r^2}{\,\(\!1\!-\!\frac{\,\Lambda r^2}{3}\!\)\,}\!
- r^2\!\(\d \theta^2\!+\sin^2\!\theta\,\d\phi^2\)\!
\\[2mm]
&=\,& \dis\d t^{'2}
\!-\!\frac{\d r^2}{\,\(\!1\!-\!\frac{\Lambda\,r^2}{3}\!\)\,}\!
-\! r^2\!\(\d \theta^2\!+\sin^2\!\theta\,\d\phi^2\) ,
\hspace*{10mm}
\ea
\eeq
where
$\,\d t =\!\(\!1\!-\!\frac{\,\Lambda r^2}{3}\!\)^{\!\!-\frac{1}{2}}\!\!\d t'$.\,
Let us make the conformal transformation,
$\,r={\overline{r}}/\!\(\!1\!+\!\frac{\Lambda}{12}\overline{r}^2\)
={\overline{r}}/\!\(\!1\!+\!\frac{1}{2}\frac{\,\over{r}^2\,}{\,r_o^2\,}\)
\simeq \over{r}$\, and
$\,{\d t'} = {\overline{R}(\overline{r})}\d\tau\,$
with
$\,\overline{R}(\overline{r})\equiv
 1/\!\(\!1\!+\!\frac{1}{2}\frac{\,\over{r}^2\,}{\,r_o^2}\)\simeq 1$,\,
where $\,\tau\,$ is the conformal time, and
the approximation sign $\,\simeq\,$ is due to
$\,r^2,\over{r}^2\ll r_o^2\,$ for any typical cosmological scale
$\,r\,$ or $\,\over{r}\,$.\,
With this transformation, we have
\beqa
\label{eq:dSS2A4e}
\dis
\d S_{\text{dS}}^2 &=&
\overline{R}^{2}(\overline{r})\!
\left[\d\tau^{2}\!- \d \overline{r}^{2}\!
-\overline{r}^2\!\(\d \theta^2\!+\sin^2\!\theta\,\d\phi^2\)\right]
=\, \overline{R}^{2}\d S_{\text{Mink}}^2
\\
&\simeq&  \d S_{\text{Mink}}^2 \,,
\nonumber
\eeqa
where $\,\d S_{\text{Mink}}^2\,$ is the usual Minkowski spacetime.
In the above, the de Sitter metric $g_{\mu\nu}^{}$ is connected to the
Minkowski metric $\eta_{\mu\nu}^{}$ via conformal transformation
$\,{g}_{\mu\nu}^{}=\overline{R}^{2}\eta_{\mu\nu}^{}\,$.\,
In the second line of Eq.\eqref{eq:dSS2A4e}, we have used the
fact that the conformal transformation factor
$\,\over{R}\simeq 1\,$ due to $\,{r}^2\ll r_o^2\,$.\,
From these, we see that the de\,Sitter spacetime is conformally flat,
and also has approximate flatness under
$\,{r}^2\ll r_o^2\,$.\,
Combining Eqs.\eqref{eq:dSS2A4c} and \eqref{eq:dSS2A4e},
we have proven that the spacetime outside the lensing region,
obeying both $\,\RRR \gg 1\,$ and $\,{r}^2/r_o^2\ll 1\,$,\,
is approximately flat.

On the other hand, the flat FLRW spacetime corresponds to
the FLRW metric with spatial curvature $K=0$\,,
\beqa
\label{eq:dSS2RW}
\dis
\d \over{S}_{\text{FLRW}}^2 =\, a^2(\tau)\!\left[\d \tau^{2} -\!\d \overline{r}^{2}
-\! \overline{r}^2\!\(\d \theta^2\!+\sin^2\!\theta\,\d\phi^2\)\right].
\hspace*{10mm}
\eeqa
Comparing Eqs.\,\eqref{eq:dSS2A4e} with \eqref{eq:dSS2RW}, we see that
 the de\,Sitter spacetime is conformally equivalent to
a flat FLRW metric with vanishing spatial curvature ($K=0$),\footnote{%
The equivalence between the de\,Sitter metric \eqref{eq:dSS2A4c}
and the FLRW metric \eqref{eq:dSS2RW} (with $K=0$) under conformal transformation
was known before and discussed in Ref.\,\cite{Weinberg08}.}
\beqa
\label{eq:FLRWflat-dS}
\d \over{S}_{\text{FLRW}}^2  \,=\,
[a(\tau )/\over{R}(\over{r})]^2\,
\d S_{\text{dS}}^2 \,\simeq\, a^2(\tau )\d S_{\text{dS}}^2 \,.
\eeqa
This shows that the de\,Sitter spacetime can be conformally and isometrically embedded into
the flat FLRW spacetime.
Hence, for $\,\RRR \gg 1\,$,\, 
the SdS metric is conformally equivalent to a flat FLRW metric
which has no effect on the deflection because of its flatness
(Bartelmann and Schneider 2001\cite{Bartelmann99}).
(We will further extend the above analysis to the case of $w\neq -1$ in Appendix\,\ref{app:B2}.)
From these, the deflection region,
$\,r\lesssim \reff \!=\!O(\rc)$,\,
is well isolated from the flat FLRW spacetime,
where $\,\rc\!\simeq \over{r}_{\text{cri}}^{}$.\,
For the transition region with $\,r\sim\rc\,$,\,
Eqs.\eqref{eq:dS2}-\eqref{eq:Phi} and \eqref{eq:Phi-CC} approximately hold.
So, the metric outside the lensing region is fairly
conformally flat, and has negligible effect on the light deflection.
For an isolated lensing system, the lensing region of
$\,r\lesssim\rc\,$ is well approximated by Eq.\eqref{eq:metric1}
where the cosmological expansion effect becomes negligible so that
$\,a(\tau)\simeq$\,constant. 
In fact, for the analyses at astrophysical scales (such as galaxies or galaxy clusters),
it is well justified to treat $\,a(\tau)\simeq$\,constant
for the data fitting\,\cite{Malin2014}\cite{Shu2016},
as we will further explain in Appendix\,\ref{app:C}.
Comparing Eq.\eqref{eq:dSS2A4c} with Eq.\eqref{eq:FLRWflat-dS},
we see that at large scales with $\,\rrr\gg\rccc\,$,\,
the original SdS metric \eqref{eq:dSS2A4} becomes approximately equivalent to
the FLRW metric \eqref{eq:dSS2RW} with $K=0$\,.\,
Hence, the deflection angle \eqref{alphadm2}-\eqref{eq:alphaDE-CC} (generated
in the lensing system $\,r\lesssim \reff\!\sim\!\rc\,$) will not be affected
when the light ray travels to the faraway region $\,\rrr\gg\rccc\,$.\,
This is expected because the regions with approximately flat spacetime
do not cause visible light deflection.
Also, any conformal transformation
[such as \eqref{eq:FLRWflat-dS} or \eqref{eq:dSS2A4e}]
does retain the angles unchanged.

We note that the conventional lensing analyses\,\cite{Bartelmann99} choose the FLRW metric
with matter density treated as small perturbation.
The observational data show that the FLRW metric is nearly flat ($K\simeq 0$),
while the matter perturbation only generates the Newtonian deflection.
So, for the region with $\,\RRR\gg 1\,$,\, it is clear that
the light bending effect of the cosmological constant $\Lambda$
is either absent or negligibly small due to the flatness of the FLRW metric,
as often found in the literature\,\cite{Ishak-Rev}.
On the other hand, people usually restrict the deflection region
to be nearby the immediate neighborhood of the matter distribution
(with $\,r\ll\rc\,$), rather than the effective region
$\,0.3\rc < r < \reff\!\sim\!\rc\,$
(as we studied in Sec.\,\ref{sec:4} and Fig.\,\ref{fig:4}).
This is another reason that some literature\,\cite{Ishak-Rev} found the $\Lambda$ bending effect
to be vanishing or negligible nearby the matter distribution.
The similar point was also clarified before by Ishak, Rindler and Dossett\,\cite{Ishak2010}.

\vspace*{1.5mm}

Before concluding this Appendix, we note that the lensing effect of
$\Lambda$ on light bending was considered before via different approaches as
reviewed by Ishak and Rindler\,\cite{Ishak-Rev}, who clarified some debates in the
literature\,\cite{Ishak2007}\cite{Ishak2010}\cite{Arakida12}\cite{1983,Ishak2008,Italy,other1}.
A main cause of the debates is the lack of a consistent approach
which could treat the lensing system as a well-defined isolated system
{\it characterized by its critical radius} $\,\rc\,$
(rather than by the radius of the matter distribution of the lens).
It was not well realized that within this isolated system, the matter potential and
dark energy potential should be treated on equal footing and solved together
from the Einstein equation under SdS metric
(rather than the nearly flat FLRW metric of large cosmological scales);
while for regions outside this system with $\,\rrr\!/\rccc\gg 1\,$,\,
the spacetime reduces to the de\,Sitter metric and
conformally recovers the FLRW metric (with $K=0$).
Our approach provides an independent and conceptually clean resolution to this issue,
which also favors Ishak and Rindler\,\cite{Ishak-Rev}.

In the following, we give some further clarifications on the literature,
making it clear that these {do not} affect our
independent and self-contained approach.~(i).~Islam\,\cite{1983}
noticed that $\Lambda$ does not modify
the light orbital equation \eqref{eq:lorbit2CC} (a second order ODE)
and thought that $\Lambda$ does not affect light bending.
Then, Rindler and Ishak\,\cite{Ishak2007} first found
that $\Lambda$ can still generate light bending via the metric itself
although Eq.\eqref{eq:lorbit2CC} does not depend on $\Lambda$\,.~Ref.\,\cite{Arakida12}
further noted that
the light orbital equation in its form of the first-order ODE \eqref{eq:lorbit1CC}
does contain an explicit $\Lambda$ term, but it could be absorbed into the definition
of an effective impact parameter $\,B\,$ so that \eqref{eq:lorbit1CC} takes
the form \eqref{eq:lorbit1CC2}.
This seems to make the $\Lambda$ effect not directly testable.
They also made the thin lens approximation to
absorb the $\Lambda$ effect into angular diameter distance via
$\,B\approx D_{C\mathcal{O}}^{}\theta\,$.\,
But this thin lens relation does not generally hold
when $\theta_c^{}$ is significant (cf.\ our Fig.\,\ref{fig:2}).
We point out that $B$ could make a sense as an effective impact parameter
only for the special case of $\,w=-1\,$ {\it and}\, point-like light source.
For all dark energy models with $\,w\neq -1\,$,\, the $w$-dependent term
in Eq.\eqref{eq:lorbit1} or \eqref{eq:1stOE-B} is $r$-dependent and a universal
effective impact parameter $B$ does not exist.
As we clarified below Eq.\eqref{eq:lorbit1CC2},
it holds only for point-like light source,
but not the more realistic 2d sources.
The impact parameter $\,b\,$ of such 2d sources varies its value
on the surface of the source. Hence, no universal effective impact parameter $B$ exists
to fully absorb $\Lambda$ effect, and
fitting the lensing data from 2d light sources can
discriminate the $\Lambda$-induced deflection effect.~(ii).~Ishak\,\cite{Ishak2008}
also analyzed the $\Lambda$ contribution to light bending
and time delays from integrating the potential term as well as from the
Fermat's principle. The pure de\,Sitter metric was used for calculations
inside the vacuole (the effective lensing area).
This appears not well justified since the pure de\,Sitter metric is conformally flat
[cf.\ our Eq.\eqref{eq:dSS2A4e}],
and thus does not cause light bending.~(iii).~Under the weak deflection approximation, Refs.\,\cite{Ishak-Rev}\cite{Ishak2007}
found the $\Lambda$-induced light bending,
$\,\Delta\alpha_{\Lambda}^{}=-\frac{\,{\Lambda}R^3}{12M\,}\,$,\,
which diverges as $M\to 0$\, and appears inconsistent
with the weak deflection assumption.
As we clarified at the end of Sec.\,\ref{sec:3} and
in the paragraph below Eq.\eqref{eq:alphaDE-CC},
the deflection $\,\Delta\alpha_{\DE}^{}$\,
consistently vanishes in the limit $\,M\to 0\,$,\, as expected.
(iv).~In the lensing system of $\,r \lesssim \rc\,$,\,
the matter contribution is larger than that of dark energy,
and the matter distribution has significant structures.
Hence, the cosmological principle no longer holds at such scales
and here the FLRW metric is not a valid description of the spacetime.
This means that for $\,r \lesssim \rc\,$,\,
it is not justified to use FLRW metric with matter density
treated as small perturbation.
As shown in Appendices\,\ref{app:11} and \ref{app:44},
the spacetime geometry in the region $\,r \lesssim \rc\,$ is best described
by the metric \eqref{eq:dSS2} [or \eqref{eq:metric1}] with the
gravitational potential \eqref{eq:Phi-NDE2} [or \eqref{eq:Phi}],
which reduces to the SdS metric \eqref{eq:dSS2A4}
with dark energy potential \eqref{eq:Phi-CC}
for the cosmological constant case
($w=-1$).~(v).\,Simpson\,{\it et al.}~in \cite{other1}
defined the vacuole boundary of lensing system by requiring
the potential $\,\Phi =0\,$ at $\,r=R\,$ below their Eq.(15).
Then, they derived light bending angle
$\,\alpha = \alpha_{\M}^{}+\alpha_{\Lambda}^{}=
(4M/R - \Lambda R r_b^{}/3)\sqrt{1\!-\!R^2/r_b^2\,}\,$
in their Eq.(32), where both $\,\alpha_{\M}^{}=0\,$ and
$\,\alpha_{\Lambda}^{}=0\,$ hold at the boundary $\,R=r_b^{}\,$.\,
This appears inconsistent because it even makes the conventional
matter-induced light bending $\,\alpha_{\M}^{}\,$
(\`{a} la Einstein\,\cite{Weinberg72}) vanish.
Moreover, imposing the condition $\,\Phi =0\,$ at the
vacuole boundary appears improper because
our gravitational potential \eqref{eq:Phi} (derived from the Einstein
equation in Appendix\,\ref{app:11}) proves that
both the Newtonian and dark energy terms
always share the {\it same sign} and cannot cancel each other to give
$\,\Phi =0\,$.\,
In fact, as we showed in Eq.\eqref{eq:force} and Eq.\eqref{eq:rc}
(including the special case $\,w=-1\,$),
the Newtonian force and the dark energy force
(rather than their potential terms)
exactly cancel at the critical radius $\,\rc\,$.\,
Hence, the boundary of an isolated lensing system
should be correctly characterized by the effective radius around its
critical radius, $\,\reff\sim\rc\,$.
(vi).~In the literature\,\cite{other1},
it is sometimes argued that
the cosmological comoving observer has a local Hubble velocity
$\,v_L^{}\simeq H\,D_L^{}\,$
(with $D_L^{}$ the luminosity distance from the observor),\, 
which causes an aberration factor to the deflection.
In fact, the Hubble velocity is not a real physical velocity,
and does not apply to the aberration equation.
It is just an apparent velocity and is solely determined
by the expansion factor $a(t)$,
so it is a cosmological expansion effect.
The factor $a(t)$ causes no visible effect on the deflection, 
especially the ratio $\,\Delta{\alpha}_{\DE}^{}/\alpha_{\M}^{}$
\cite{Ishak2010}, 
because the expansion is {\it conformally flat} for $\,K\simeq 0\,$,\,
as discussed around Eqs.\eqref{eq:dSS2A4c}-\eqref{eq:FLRWflat-dS}.


\vspace*{3mm}
\section{\hspace*{-1.5mm}Deriving Light Orbital Equation for Generic State Parameter w}
\vspace*{1mm}
\label{app:B}
\label{app:B1}

In this Appendix, we will derive the new light orbital equation \eqref{eq:lorbit1} for
generic state parameter $w$\, and under the SdS$w$ metric.

For a gravitational lensing system with generical dark energy state parameter $w$,\,
we have the SdS$w$ spacetime metric from Eq.\eqref{eq:dSS2},
\beqa
\label{eq:dSS2A5}
\dis
\d S^2 = \left[1\!-2\frac{M}{r}\!-2\!\left(\!\!\frac{\,\,\ro\,}{r} \!\!\right)^{\!3w+1} \right]\!\d t^{2}
-\frac{\d r^2}{\,\left[1\!-2\frac{M}{r}\!-2\!\left(\!\frac{\,\ro\,}{r} \!\right)^{\!3w+1}\right]\,}
- r^2\!\(\d \theta^2\!+\sin^2\!\theta\,\d\phi^2\)\!.
\hspace*{10mm}
\eeqa

Let us define the energy-momentum four-vector
$\,K^{\mu}={\,\d x^{\mu}}/{\d \lambda}\,$,\,
where the $\,\lambda\,$ is an affine parameter.
From Eq.\eqref{eq:dSS2A5}, we have
\beqa
g_{00}^{}=\!1\!-2\frac{M}{r}\!-2\!\left(\! \frac{\,\ro\,}{r} \!\right)^{\!3w+1},~~~~
g_{11}^{}= -g_{00}^{-1},~~~~ g_{22}^{} = -r^2,~~~~
g_{33}^{}= -r^2\sin^2\!\theta \,,~~~~~
\eeqa
and $\,g_{ij}^{}=0\,$ for $\,i\neq\,j$\,.\,
Since this metric respects the symmetries of time translation and space rotation,
it holds the conservations of energy $E$ and angular momentum $L$,\, i.e.,
$\,E=$\,constant and $\,L=$\,constant. So we have,
\beqa
\label{eq:enanEqC5a}
\ba{lcl}
\dis
E &=& \dis
g_{0\mu}^{}\,K^{\mu}
= \left[1\!-2\frac{M}{r}-2\!\(\! \frac{\,\ro\,\,}{r} \!\!\)^{\!3w+1} \right]
\frac{\d t}{\d \lambda} \,,
\\[4mm]
L &=& \dis g_{3\mu}^{}\,K^{\mu} = -r^2\sin^2\!\theta\frac{\d \phi}{\,\d \lambda\,} \,.
\hspace*{10mm}
\ea
\eeqa

Using the null condition $\d S^2=0$\, for a light ray, we have
\beqa
\label{eq:nEqC5b}
\dis
\left[1\!-2\frac{M}{r}-2\!\(\! \frac{\,\ro\,\,}{r}\!\!\)^{\!3w+1} \right]\!\d t^{2}
-\frac{\d r^2}{\,\left[1\!-2\frac{M}{r}-2\!\left(\! \frac{\,\ro\,}{r} \!\right)^{\!3w+1}\right]\,}
- r^2\!\(\d \theta^2\!+\sin^2\!\theta\,\d\phi^2\) =\, 0 \,.
\hspace*{10mm}
\eeqa
Without losing generality,
we confine the motion in the plane of $\,\theta =\frac{\pi}{2}$.\,
Thus, we can rewrite Eqs.\eqref{eq:enanEqC5a} as follows,
\beq
\label{eq:nEqC}
\ba{rcl}
\dis
\frac{\d t}{\d \lambda} &=&
\dis
E\left[1\!-2\frac{M}{r}-2\!\(\! \frac{\,\ro\,}{r} \!\)^{\!3w+1}\right]^{-1} \!,
\\[4mm]
\dis \frac{\,\d \phi\,}{\d \lambda} &=& \dis -\frac{L}{\,r^2\,} \,.
\ea
\eeq
With these, we further rederive Eq.\eqref{eq:nEqC5b} as
\beqa
\label{eq:dS2=0}
\dis\(\!\frac{\d r}{\,\d \lambda\,}\!\)^{\!\!2}
&=& \dis E^2-\frac{\,L^2}{r^2}\!\left[1\!-2\frac{M}{r}-2\!\(\! \frac{\,\ro\,}{r} \!\)^{\!3w+1}
\right]\!.
\hspace*{10mm}
\eeqa
Substituting the second formula of Eq.\eqref{eq:nEqC} into Eq.\eqref{eq:dS2=0},
we deduce the following light orbital equation for general state parameter $w$\,,
\beqa
\label{eq:lightLC5}
\dis
\(\!\frac{1}{r^2}\frac{\d r}{\d \phi}\!\)^{\!\!2}
=\, \frac{\,E^2\,}{\,L^2\,}-\frac{1}{\,r^2\,}\!
\left[1\!-2\frac{M}{r}-2\!\left(\! \frac{\,\ro\,}{r}\!\right)^{\!3w+1}\right]
\hspace*{10mm}
\eeqa
For convenience, let us define the notations $\,u={1}/{r}$\, and $\,b=L/E$\,.\,
Hence, we can reexpress Eq.\eqref{eq:lightLC5} as follows,
\beqa
\label{eq:lorbit2}
\left(\frac{\d u}{\d\phi}\right)^{\!\!2}
=~\frac{1}{\,b^2\,}-u^2+2Mu^3+2\,{\ro}^{\!\!3w+1}u^{\!3(w+1)} \,,
\hspace*{9mm}
\eeqa
which just reproduces the generical light orbital equation \eqref{eq:lorbit1}
which we presented in Sec.\,\ref{sec:55}.

\vspace*{3mm}
\section{\hspace*{-1.5mm}Analyses within and outside the Lensing System}	
\label{app:CC}
\vspace*{1mm}

In this Appendix, we first show that the cosmic expansion effect is negligible
for a lensing system at typical astrophysical scales.
Then, we will prove the approximate conformal flatness of the SdS$w$ metric
in the region $\,r^3/\rccc\gg 1\,$.\,

\vspace*{2mm}
\subsection{\hspace*{-1.5mm}Cosmic Expansion Factor is Nearly Constant at Astrophysical Scales}
\label{app:C}
\label{app:C1}


In the following, we show that
for a lensing system at typical astrophysical scales (such as galaxies or galaxy clusters),
the cosmic expansion factor is nearly constant.
Although this is a known fact, it useful to explicitly clarify it
here for supporting the current formulation.

For a given astrophysical object, its luminosity distance from us as the observer is denoted as $D_L^{}$,
and its size is denoted as  $\Delta D_L^{}$.
According to the Hubble Law, we have
\beqa
\label{eq:hubblelaw1}
z\,\simeq\,v_L^{} =\, H\,D_L^{}\,,
\eeqa
where the $v_L^{}$ is the velocity along the sight direction, the $\,z\,$ is the cosmological redshift.
Then, we also have
\beqa
\label{eq:dhubblelaw2}
\Delta z \,\simeq\, H\,\Delta D_L^{}\,,
\eeqa
where the $\Delta z$ describes the redshift fluctuation due to the finite size of this astrophysical object.
Combining Eqs.\eqref{eq:hubblelaw1} and \eqref{eq:dhubblelaw2}, we deduce
\beqa
\label{eq:dZvsZ}
\frac{\,\Delta z\,}{z} \simeq \frac{\,\Delta D_L^{}\,}{\,D_L^{}\,}.
\eeqa
Let $a=a(\tau)$ be the expansion factor at redshift $z$\,,\,
and $\,a_0^{}\!=a(\tau_0^{})$\, be the expansion factor at $z=0$\,.\,
Thus, we have
\beqa
\label{eq:factor1}
\frac{a}{\,a_0^{}\,} \,=\, \frac{1}{\,1\!+\!z\,} \,.
\eeqa
The redshift fluctuation $\,\Delta z\,$ in Eq.\eqref{eq:dhubblelaw2}
corresponds to a variation $\Delta a$ of the cosmic expansion factor $\,a\,$.\,
Including this effect, we reexpress the formula \eqref{eq:factor1},
\beqa
\label{eq:factor2}
\frac{\,a\!+\!\Delta a\,}{a_0^{}} \,=\,
\frac{1}{\,1\!+\!z\!+\! \Delta z\,} \,.
\eeqa
With Eqs.\,\eqref{eq:factor1} and \eqref{eq:factor2}, we deduce
\beqa
\label{eq:davsa}
\frac{\,|\Delta a|\,}{a} \,=\, \frac{\,|\Delta z|\,}{1\!+\!z}
\,\simeq\, \frac{\,|\Delta D_L^{}\!|\,}{D_L^{}}
\frac{z}{\,1\!+\!z\,} \,.
\eeqa

The galaxies and galaxy clusters are the gravitationally self-bounded systems
in the Universe, with typical mass range $\,M \!=\! (10^{12}\!-\!10^{16})M_{\odot}^{}$.\,
In Sec.\,\ref{sec:4}, we estimated the range of their critical radii,
$\,\rc\simeq (1.1-23)$\,Mpc, for $\,w= -1$\,.\,
Thus, we can expect ${\Delta D_L^{}}\lesssim \rc \simeq (1.1-23)$\,Mpc.
For instance, we consider the current lensing experiments
for galaxies with masses $\,M \lesssim 10^{13}M_{\odot}^{}$\,
and at redshift $\,z=0.1-0.5$ \cite{Malin2014,Shu2016}.
We can estimate $\,D_L^{}=z/H \simeq (0.42-2.1)\!\times\!10^3$\,Mpc
and $\,{\Delta D_L^{}}\lesssim\rc \lesssim 2.3$\,Mpc.
Thus, we have
$\,\frac{|\Delta z|}{z} \lesssim (5.5-1.1)\!\times\!10^{-3}$
and $\,\frac{|\Delta a|}{a} \lesssim (5.0-3.7)\!\times\!10^{-4}\ll 1$\,.\,
Then, we consider the galaxies or galaxy clusters with masses
$\,M=10^{12-16}M_{\odot}^{}$\,
and at redshift $\,z=0.5-2$ \cite{Zitrin11,Sluse12,Umetsu16}.\,
Thus, $\,D_L^{}=z/H \simeq (2.1\!-\!8.4)\!\times\!10^3$\,Mpc for $\,z=0.5-2$,
and $\,{\Delta D_L^{}}\lesssim\rc \simeq (1.1-23)$ Mpc for
$\,M=10^{12-16}M_{\odot}^{}$.\,
Thus, we can estimate $\,\frac{|\Delta z|}{z}\lesssim (0.52\!-\!11)\!\times\!10^{-3}$
and $\,\frac{|\Delta a|}{a} \lesssim (0.17\!-\!3.7)\!\times\!10^{-3}\ll 1$\,
at $\,z=0.5$;\,
while $\,\frac{|\Delta z|}{z}\lesssim (0.13\!-\!2.7)\!\times\!10^{-3}$
and $\,\frac{|\Delta a|}{a} \lesssim (0.08\!-\!1.8)\!\times\!10^{-3}\ll 1$\,
at $\,z=2$.\, The above analysis shows that for a typical galaxy or galaxy cluster
as the gravitational lensing, the cosmic expansion effect is negligible
and the lensing system is well described by the static SdS$w$ spacetime
\eqref{eq:dSS2}-\eqref{eq:Phi-NDE2}.

\vspace*{2mm}
\subsection{\hspace*{-1.5mm}Approximate Conformal Flatness for Generic State Parameter w}
\vspace*{1mm}
\label{app:B2}
\label{app:C2}

In this Appendix, we show that for the region outside the lensing system with $\,\RRR \gg 1\,$,\,
the general SdS$w$ metric \eqref{eq:dSS2}-\eqref{eq:Phi-NDE2} has approximate conformal flatness.
This is an extension of our discussion in Appendix\,\ref{app:A4}.
Since $w$ is quite close to $\,w=-1$ \cite{BAO}, we see that the condition
$\,\RRR \gg 1\,$ lead to
$\,\Phi_{\N}^{}/\Phi_{\DE}^{}\sim r_{\text{cri}}^{3\left|w\right|}/r^{3\left|w\right|} \ll 1\,$.\,
%
%
Hence, the Newtonian potential $\Phi_{\N}^{}$
becomes negligible for $\,\RRR \gg 1\,$,\, and thus the SdS$w$ metric
\eqref{eq:dSS2}-\eqref{eq:Phi-NDE2} approximately reduces to the
de Sitter spacetime with generical $w$ (denoted as dS$w$),
\beqa
\label{eq:dSS2A5c}
\d S^2 \simeq\,\d S_{\text{dS}w}^2
\dis
&=& \left[\!1\!-2\!\left(\! \frac{\,\ro\,}{r} \!\right)^{\!3w+1} \!\right]\!\d t^{2}
\!- \frac{\d r^2}{\left[\!1\!-2\!\left(\! \frac{\,\ro\,}{r} \!\right)^{\!3w+1} \!\right]}\!
- r^2\!\(\d \theta^2\!+\sin^2\!\theta\,\d\phi^2\)\!
\nn\\[2mm]
&=& \d t^{'2}
\!- \frac{\d r^2}{\left[\!1\!-2\!\left(\! \frac{\,\ro\,}{r} \!\right)^{\!3w+1} \!\right]}\!
- r^2\!\(\d \theta^2\!+\sin^2\!\theta\,\d\phi^2\)\!,
\hspace*{10mm}
\eeqa
where
$\,\d t =\!
 \left[\!1\!-2\!\left(\! \frac{\,\ro\,}{r} \!\right)^{\!3w+1} \!\right]^{\!-\frac{1}{2}}\!\!\d t'$.\,
Then, we make the coordinate transformations,
$\,r={\overline{R}(\overline{r})}\,{\overline{r}}$\, and
$\,{\d t'} = {\overline{R}(\overline{r})}\,\d\tau\,$,\,
where $\,\tau\,$ is the conformal time and ${\overline{R}(\overline{r})}$ obeys the condition,
\beqa
\label{eq:Rbar-Cond}
\frac{\,\d \ln(\rh\RB )\,}{\d\ln\rh} \,=\,
\left[1-2\!\(\rh\RB\,\)^{\!-(3w+1)}\right]^{\!\frac{1}{2}}\!,
\eeqa
where $\,\rh \equiv \rb/\ro\,$.\,
From Eq.\eqref{eq:Rbar-Cond}, we derive the solution
\beqa
\label{eq:Rbar-w}
\RB (\rb) &=& \frac{\,\ro\,}{\,\rb\,}\!\left[
\frac{1}{2}\sin^2\!\(\!2\arctan\!\frac{\,\,(\rb/\ro)^{\frac{|3w+1|}{2}}}{\sqrt{2}}\!\)\!
\right]^{\!\frac{1}{|3w+1|}}
\nn\\
\\[-5.5mm]
&\simeq& 1- \frac{~(\rb /\ro)^{|3w+1|}\,}{|3w+1|}
\,\simeq\, 1 \,,
\hspace*{10mm}
(\text{for}~ \rb^2 \ll r_o^2\,),
\nn
\eeqa
where $\,3w+1<0\,$ always holds due to the current data\,\cite{BAO}.
For the special case of $\,w=-1$,\, the formula \eqref{eq:Rbar-w} reduces to
$\,\overline{R}(\overline{r})=
 1/\!\left[1\!+\!\frac{1}{2}({\over{r}^2}\!/{r_o^2})\right]$,\,
which agrees to what shown below Eq.\eqref{eq:dSS2A4c}.

Under the above coordinate transformations $(t',\,r)\!\to\!(\tau,\,\rb )$,\,
we can reexpress the metric \eqref{eq:dSS2A5c} as follows
\beqa
\label{eq:dSS2A5e}
\dis
\d S_{\text{dS}w}^2 &=&
\overline{R}^{2}(\overline{r})\!
\left[\d\tau^{2}\!- \d \overline{r}^{2}\!
-\overline{r}^2\!\(\d \theta^2\!+\sin^2\!\theta\,\d\phi^2\)\right]
=\, \overline{R}^{2}\d S_{\text{Mink}}^2,
\eeqa
which means
\beqa
\label{eq:FLRWflat-w}
\d \over{S}_{\text{FLRW}}^2  \,=\,
[a(\tau )/\over{R}(\over{r})]^2 \d S_{\text{dS}w}^2\,.
\eeqa
This shows that the dS$w$ metric is equivalent to the flat FLRW metric
($K=0$) under conformal transformation.
Hence, from these we see that in the region $\,\RRR \gg 1\,$,\,
the SdS$w$ metric becomes conformally equivalent to a flat FLRW metric ($K=0$),
and thus causes no visible light deflection.
The above is an extension of our discussion around
Eqs.\eqref{eq:dSS2A4}-\eqref{eq:FLRWflat-dS} of Appendix\,\ref{app:44}
to the case with a generic state parameter $w$\,.\,

\end{appendix}

\baselineskip 17pt

\vspace{3mm}
%

\end{document}